\documentclass[3p,number,sort&compress,times,12pt]{elsarticle}

\usepackage{graphicx}

\usepackage{amssymb,amsmath}
\usepackage[T1]{fontenc}
\usepackage{courier}
\usepackage{tabularx}
\usepackage{enumitem}

\journal{Computer Physics Communications}

\begin{document}

\begin{frontmatter}

\title{SHARE with CHARM\tnoteref{support}}

\tnotetext[support]{Work supported by  
the U.S. Department of Energy, grants DE-FG02-04ER41318 (MP, JL, JR) and DE-FG02-93ER40764 (GT). Laboratoire de Physique Th{\' e}orique et Hautes Energies, LPTHE, at University Paris 6 is supported by CNRS as Unit{\' e} Mixte de Recherche, UMR7589. GT acknowledges  financial support received from the Helmholtz International Centre for FAIR within the framework of the LOEWE program (Landesoffensive zur Entwicklung Wissenschaftlich-\"{o}konomischer Exzellenz) launched by the State of Hesse.}

\author[arizona,ctu]{M.~Petran}

\author[arizona,paris]{J.~Letessier}
\author[arizona]{J.~Rafelski}
\author[GT1,GT2]{G.~Torrieri}

\address[arizona]{Department of Physics, The University of Arizona, Tucson, AZ 85721, USA}%
\address[ctu]{Czech Technical Univesity in Prague, Czech Republic}%
\address[paris]{Laboratoire de Physique Th\'{e}orique et Hautes Energies, Universit\'{e} Paris 6, Paris 75005, France}%
\address[GT1]{FIAS, J.W.~Goethe Universit\"{a}t, 60438 Frankfurt A.M., Germany}%
\address[GT2]{Pupin Physics Laboratory, Columbia University, 538 West 120$^{th}$ Street, NY 10027, USA}%

\begin{abstract}
SHARE with CHARM  program (SHAREv3) implements the statistical hadronization model description of particle production in relativistic heavy-ion collisions. Given a set of statistical parameters, SHAREv3  program evaluates yields and therefore also ratios, and furthermore, statistical particle abundance fluctuations. The physical bulk properties of the particle source is evaluated based on all hadrons produced, including the fitted yields. The bulk properties can be prescribed as a fit input complementing and/or replacing the statistical parameters. The modifications and improvements in the SHARE suite of programs are oriented towards recent and forthcoming LHC hadron production results including charm hadrons. This SHAREv3 release incorporates all features seen previously in SHAREv1.x and v2.x and, beyond, we include a complete treatment of charm hadrons and their decays, which further cascade and feed lighter hadron yields. This article is a complete and self-contained manual explaining and introducing both the conventional and the extended capabilities of SHARE with CHARM. We complement the particle list derived from the Particle Data Group tabulation~\cite{Beringer:1900zz-front} composed of up, down, strange  $u,d,s$ quarks (including resonances) with hadrons containing charm $c,\bar c$ quarks. We provide a table of the charm hadron decays including partial widths. The branching ratios of each charm hadron decays add to unity, which is achieved by including some charm hadron decay channels based on theoretical consideration in the absence of direct experimental information. A very successful interpretation of all available LHC results has been already obtained using this program.
\end{abstract}

\begin{keyword}
statistical hadronization model\sep  SHM, quark-gluon plasma\sep QGP\sep strangeness production\sep charm production\sep hadron  fluctuations\sep  relativistic heavy-ion collisions\sep RHIC\sep LHC


\end{keyword}

\end{frontmatter}


{\bf NEW VERSION PROGRAM SUMMARY}

\begin{small}
\noindent
{\em Program Title:} SHARE with CHARM\\
{\em Journal Reference:}                                      \\
{\em Catalogue identifier:}                                   \\
{\em Licensing provisions:} none                              \\
{\em Programming language:} FORTRAN77, C++                    \\
{\em Computer:} PC, Intel 64-bit, 3~GB RAM (not hardware dependent)\\
{\em Operating system:} GNU Linux: Ubuntu, Debian, Fedora (not OS dependent)\\
{\em RAM:} 615 MB                                              \\
{\em Number of processors used:} 1 (single thread)                              \\
{\em Keywords:} SHARE, statistical hadronization model,  SHM, quark-gluon plasma, QGP, strangeness production, charm production, hadron  fluctuations,  relativistic heavy-ion collisions, RHIC, LHC \\
{\em Classification:} 11.2, 11.3               \\
{\em External routines/libraries:}  Standard C++ library, CERNLIB library\\
{\em Catalogue identifier of previous version:} ADVD\textunderscore v2\textunderscore 0      \\
{\em Journal reference of previous version:} Comput.Phys.Commun. {\bf 175} (2006) 635-649   \\
{\em Does the new version supersede the previous version?:} Yes \\
{\em Nature of problem:}
The Understanding of  hadron production incorporating the  four $u,d,s,c$ quark flavors is essential for the  understanding of  the properties of quark--gluon plasma created in relativistic heavy-ion collisions in the  large-hadron collider (LHC) energy domain. We describe  hadron production by a hot fireball within  the statistical hadronization model (SHM) allowing for the chemical nonequilibrium of all quark flavors individually.  By fitting particle abundances  subject to bulk property constraints in the source, we find the best SHM model parameters. This approach allows  to test  physical hypotheses regarding hadron production mechanisms in relativistic heavy-ion collisions,  physical properties of the source at hadronization and the validity of the statistical hadronization model itself. The abundance of light hadrons made of $u$, $d$ and $s$ constituent quarks~\citep{Torrieri:2004zz-front} and their fluctuations~\citep{Torrieri:2006xi-front} were the core physics contents of the prior releases  SHAREv1.x and v2.x respectively. We now consider the hadronization of the heavier charm quarks, a phenomenon of relevance in the analysis of recent and forthcoming LHC results. We introduce  bulk matter constraints such as a prescribed charge to baryon ratio originating in the initial state valance $u$ and $d$ quark content of colliding nuclei. More generally, all the  bulk physical properties of the particle source  such as energy, entropy, pressure, strangeness content and baryon number of the fireball at hadronization are evaluated and all of these can be used as fit constraints. The charm quark degree of freedom is handled as follows: given an input number of charm quark pairs at the time of charm chemical freeze-out, we  populate charm hadron yield according to rules of statistical hadronization for a prescribed set of  parameters associated with the particle source, such as  bulk matter fugacities.  A seperate charm hadronization temperature can be chosen and fitted, and as an option it is possible to make this temperature the same as the fitted hadronization temperature of $u,d,s$-quarks.  Charm hadron  resonances decay feeding `stable' charm hadrons. These stable charm hadrons are so short-lived that within current technological detector capabilities practically all their decay products are  feeding  light hadron yields. These charm decay feeds are changing the abundances of produced hadrons in a pattern  that differs from particle to particle. \\
{\em Solution method:}
SHARE with CHARM builds in its approach upon the numerical method developed for its predecessor, SHARE~\citep{Torrieri:2004zz-front,Torrieri:2006xi-front} for the  evaluation of the distribution of light ($u,d,s$) hadrons. SHARE with CHARM   distributes a prescribed number $N_{c\bar{c}}$ of charm $c+\bar c$ quarks into individual charm hadrons   applying statistical hadronization rules in a newly added computation module `CHARM'  obtaining the yields evaluating appropriate  series of Bessel functions. Similarly to light hadrons, the charm hadrons decays are evaluated using pre-existent tables derived from PDG listing~\cite{Beringer:1900zz-front}, proceeding from the heaviest to the lightest particle. The  yields of each hadron are obtained  using decay branching ratio tables of the mother particle yield -- where data was not available, appropriate theoretical model was implemented to assure that all particles decayed with 100\% probability. Each of the resultant daughter hadron contributions is added to this  $u,d,s$ hadron yield computed independently for the related set of SHM parameters in the SHARE module. The total yield is subsequently subject to  the  weak decays (WD) of strange hadrons. A user generated  or default WD control file  defines what portion of the $u,d,s$ particle yield decays weekly feeding other particles in turn, and which fraction  given the detection capability is observed. Once final observable hadron yields are so obtained, we compare these with the experimental data aiming in an iteration to find the best set of prescribed SHM parameters for the yield of $u,d,s$ hadrons observed.. The CHARM module is associated with two new SHM parameters, the  charm hadronization temperature $T_{charm}$ (which can be defaulted to $T$ obtained for the other $u,d,s$ hadrons) and the total yield of $N_{c\bar{c}}=c+\bar c$ quarks, called \texttt{Ncbc}. These and all other SHM parameters are discussed in text.\\
{\em Reasons for the new version:}*\\
Since the release of SHAREv1 in 2004~\cite{Torrieri:2004zz-front} and SHAREv2 in 2006~\cite{Torrieri:2006xi-front}, heavy-ion collision experiments underwent major development  in both detector technology  and collision energy. The forthcoming tracker upgrade of STAR at BNL Relativistic Heavy Ion Collider (RHIC) and the current tracking precision of ALICE at CERN Large Hadron Collider (LHC) require upgrades of the SHARE program described below. In the anticipation of significant charm abundance at LHC, SHARE with CHARM allows the study of  all charm hadron production. Charm hadron decays are particularly important because they are a significant source of multistrange hadrons. The introduction of charm component  of the hadron spectrum into SHM is crucial for correct interpretation of particle production and QGP fireball properties at hadronization in heavy-ion collisions at TeV energy scale. SHARE with CHARM is an easy-to-use program, which offers a common framework for SHM analysis of all contemporary heavy-ion collision experiments for the coming years. \\
{\em Summary of revisions:}*\\
The charm hadron mass spectrum and decays have been fully implemented in the provided program package. We provide a current up-to-date detailed list of charm hadrons and resonances together with their numerous decay channels within the set of fully updated input files that correspond to the present PDG status~\cite{Beringer:1900zz-front}. Considering the enhanced tracking capabilities of LHC experiments and similar RHIC capability, the default behavior of weak decay feed-down has been updated to not accept any weak feed-down unless specified otherwise by the user. The common framework for all contemporary heavy-ion experiments required an update of the format of the particle list and of the content to correspond to present day PDG. SHARE with CHARM is backward compatible with the previous release, SHAREv2, in terms of calculation capabilities and use of control files. However, SHARE user may need to update and or add  individual input file command lines in order to assure that same tasks are performed, considering that defaults, e.g., characterizing  weak decays, have been modified. Furthermore quite a few  interface improvements have been implemented and are described in detail further in this manual. They allow considerable simplification of control files.\\
{\em Running time:} From a few seconds in case of calculating hadron yields and bulk properties given a prescribed set of model parameters, to $\sim 30$ hours in case of fitting all parameters to experimental data and calculation with finite widths. Sample calculation provided in the program package, which demonstrates the program capabilities other than calculation with finite widths, took just under 2 hours on both 2.1 GHz CPU (2MB L2 cache) laptop and 2.5 GHz CPU (6MB L2 cache) cluster computing node. Simple fit of model parameters to a data set (provided as default in the package) takes about 5 minutes.\\

\end{small}


\section*{Quick Start reference}
\subsection*{Installation}
This section provides a quick reference how to install SHARE with CHARM on most common PC with GNU Linux system, namely we assume \texttt{gfortran} and \texttt{g++} compilers and \texttt{cernlib} installed. If you encounter any problems following this quick guide, please refer to Section~\ref{sec:installation} for detailed installation guide.
\begin{enumerate}[leftmargin=*]
\setlength{\parskip}{0pt}
\item In a terminal, navigate into a folder where you want to install SHARE with CHARM and download the package with the command\\
\verb+wget http://www.physics.arizona.edu/~gtshare/SHARE/sharev3.zip+
\item Unzip the package contents with the command (this will create a new subfolder \texttt{sharev3}) \\
\verb+unzip sharev3.zip+
\item Enter the unpacked folder using\\
\verb+cd sharev3+\\
and compile SHARE with CHARM using\\
\verb+make+
\end{enumerate}

\subsection*{Running SHARE with CHARM}
Once SHARE with CHARM is compiled, it can be run in a terminal with the command\\
\verb+./share+\\
If you have not already, it is a good idea to run the program once with the default setup. Individual operations SHARE performs during a run are specified in the file \texttt{sharerun.data}. Without any changes to the input files after installation, the program is preset to read the provided input files and to perform a chemical non-equilibrium fit to 10-20\% centrality data from Pb--Pb collisions at LHC employing only 2 free parameters $T$, $V$ -- this calculation takes typically less than 10 seconds. 

Let us show how to modify the input files in order to perform a semi-equilibrium fit to the same data set instead of the simplified full non-equilibrium. Note that for the purpose of this quick start we do not explain in full detail all inputs that will appear below.

Changing the nature of the fit requires a few steps. We begin by changing the fit output filename (so the old fit is not overwritten), than changing a parameter value,  and learning how to include the parameter among those being  fitted.
\begin{description}
\setlength{\itemsep}{-0.1cm}
\setlength{\labelwidth}{1.5cm}
\setlength{\itemindent}{0.5cm}
\item[Changing fit output file name] Looking at the contents of \texttt{sharerun.data} in a text editor, the fitting command is the following line\\
\verb+CALC  FITRATIOS  fitTESTne.out+\\
It tells SHARE to perform a fit (\texttt{FITRATIOS}) all free parameters to experimental data, and that the output file name is \texttt{fit1020ne.out}. The file is overwritten every run. So let us redirect the new output to another file by changing this line to\\
\verb+CALC  FITRATIOS  fitTESTse.out+\\
so we have both outputs for comparison. The output filename has to be 13 characters long. Remember to keep two spaces between each word.
\item[Setting parameter value] Parameter values are set in \texttt{thermo.data} file. Open it in a text editor of your choice. Chemical semi-equilibrium is defined by $\gamma_q=1$ and thus  we need to change the line starting with the parameter name, \texttt{gamq}, to read\\
\verb+gamq    1.+\\
The format is such that you must remember to keep four spaces after the parameter name and always enter a decimal point even for integer values. The values specified in this input are either fixed parameter values, or the initial fitted parameter values. Final cross check of fits with several different initial parameter values is advisable to fully understand errors and  fit stability, i.e., that the fit converges to the same minimum from different starting point(s) in the parameter space and that error is not underestimated.
\item[Fixing/Fitting a parameter] Parameter ranges for this test run defined in \texttt{ratioset.test} file (equivalent to \texttt{ratioset.data}, Section~\ref{sec:ratioset}). Upon opening the file in a text editor, you will notice that each parameter has a separate line such as the following one for \texttt{gams} ($\gamma_s$):\\
\verb+gams    0.1     9.    0.1     0+\\
In the previous non-equilibrium fit, \texttt{gams} was fixed. Parameter with \texttt{0} in the last column will be kept constant at the value specified in \texttt{thermo.data} file during a fit, whereas parameters with \texttt{1} in the last column are to be changed within the allowed range (first two numbers on the line). The different value of $\gamma_q$ we set in the previous step will result in a new value of $\gamma_s$, so let us release \texttt{gams} by changing the \texttt{0} to \texttt{1} on the above quoted line, so the line now reads:\\
\verb+gams    0.1     9.    0.1     1+\\
demanding that the program finds the best value of $\gamma_s$ to describe the data.
\end{description} 
With the above modifications, running \texttt{./share} again will produce a new output file with semi-equilibrium fit (with 3 free parameters, $T$, $V$ and $\gamma_s$) obtained for  the same experimental data defined in the file \texttt{LHC1020MI.data}. Note that the resulting fit should have lower CL  as other SHM parameters  for purpose of this example remain fixed to their optimized full non-equilibrium values. 
\begin{description}
\setlength{\itemsep}{-0.1cm}
\setlength{\labelwidth}{1.5cm}
\setlength{\itemindent}{0.5cm}
\item[Changing experimental data point] Every line in \texttt{LHC1020MI.data} contains one data point name, experimental value, statistical and systematic error and whether or not this data point is fitted or only evaluated during a fit. For example, the experimenta yield of $\Lambda=17\pm2$ is defined on the following line:\\
\verb+Lm1115zer  prt_yield     17.       2.0         0.           1+\\
One can change the value from $17\pm2$ to a different one by changing the numbers. The data point can be excluded from the fit by changing the \texttt{1} to a \texttt{0} in the last column. Similarly to fixing a parameter above, this implies that the experimental value will not be fitted, its theoretical value will be calculated based on the model parameters irrespective of its experimental value.
\end{description} 

SHARE with CHARM program is far more capable than the basic operation shown in this Quick Start guide, we refer the reader to the following 30 pages for details about program operation, input file structure, and full description of program capabilities.

\section{Physics motivation}
Strong interaction reactions usually lead to high multiplicity of produced particles. A non-perturbative description of particle production has been proposed originally by Fermi~\cite{Rafelski:2003zz} based on statistical ideas and later the model was developed further by considering the reaction volume expansion and realizing that at some point during the expansion, the particle density decreases below the point, where they can interact with each other. This stage is referred to as chemical freeze-out. Next important feature included the hadron resonance mass spectrum significantly increasing the number of states to be populated in the statistical approach. The hadron resonance spectrum implied that the hadronic matter could undergo a phase transition at Hagedorn temperature $T_H\sim 160\,\textrm{MeV}$ into a gas of quarks. For the statistical model milestones and more detailed history, see~\cite{Rafelski:2003zz} and other references in~\cite{Torrieri:2004zz}.

Relativistic heavy-ion collisions allow us to create a fireball of matter at very high temperature and density in a laboratory. The objective of the heavy-ion collision program is to study the formation of a deconfined state of matter, the quark--gluon plasma (QGP) and its transition to hadronic matter. The Early Universe has been composed of QGP up until a few microseconds after the Big Bang, when quarks and gluons merged into hadrons, particles that we see around us today. Creation of a small fireball of matter, where quarks and gluons are not bound, would confirm that deconfinement is a property of strong interaction vacuum state. An overview of the matter can be found for example in~\cite{Letessier:2002gp}.

The short lifetime and the extreme conditions leave us with indirect observations of the fireball. It is challenging to identify unique probes that allow us to distinguish between a deconfined QGP and a sequence of hadron interactions leading to the final hadron state we observe experimentally. High multiplicity of produced hadrons is a characteristic feature of heavy-ion collisions irrespective of whether or not deconfined state of matter has been achieved during the collision. Specific properties of the final hadron state can distinguish between the two scenarios of hadron production. For details about the differences in the final hadron state see, e.g.,~\cite{Koch:1986ud}.

Statistical hadronization model has been used in the past decades to describe hadron production in heavy-ion collisions at CERN Super Proton Synchrotron (SPS) ($\sqrt{s_{NN}}=8-17\,\mathrm{GeV}$), RHIC ($\sqrt{s_{NN}}=64-200\,\mathrm{GeV}$) and recently at LHC ($\sqrt{s_{NN}}=2.76\,\mathrm{TeV}$) with oftentimes great accuracy. Despite the variety of SHM approaches (chemical equilibria of different flavors, post-hadronization interactions,\ldots) has a common evolution pattern, at some point during the evolution, the phase space of stable hadrons and resonances is populated as described by their respective statistical distributions. Then, the resonances decay and thus significantly increase the yields of the daughter particles.

Proper model description of the final hadron state yields information about the source of hadrons in relativistic heavy-ion collisions and its properties at the time of hadronization, the transition from the deconfined QGP phase into hadrons. We have compiled an upgraded program `SHARE with CHARM', which produces the final hadron yields and ratios based on intensive parameters of the particle source. We have prepared a package, that takes advantage of already implemented and thoroughly tested program SHAREv2 written in Fortran 77 and we complement it with an external module written in C++ which adds proper description of charm hadron production according to the current status of the field including updated input data tables. 

\begin{figure}[t]
\centering
\includegraphics[height=110mm]{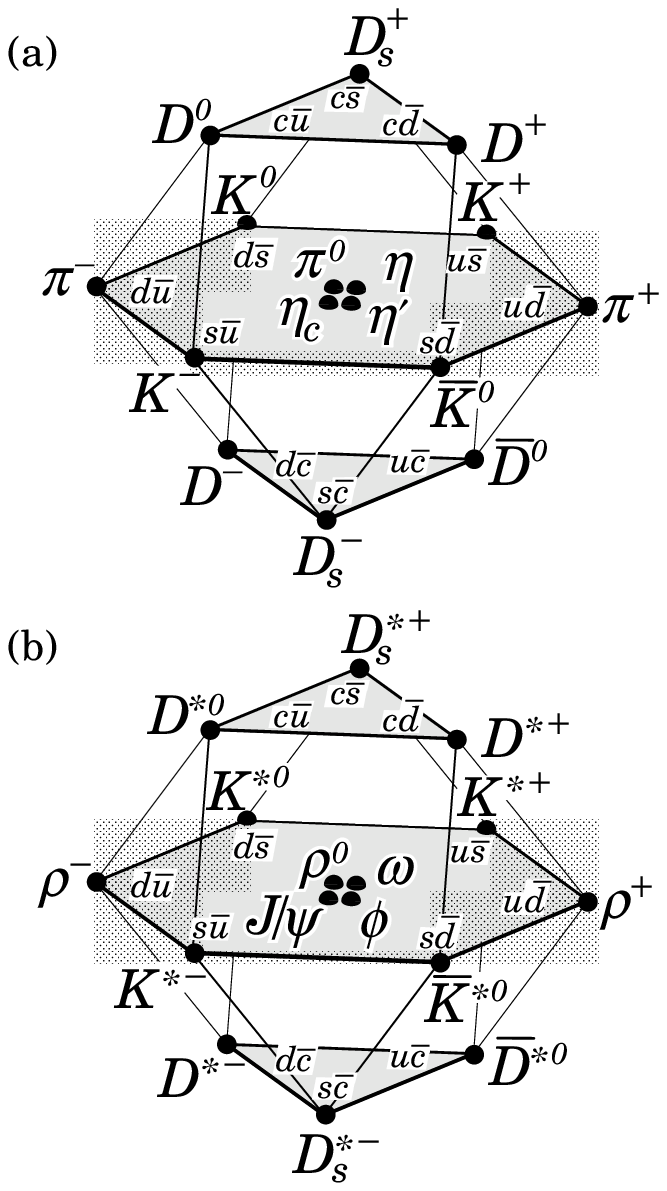}\hspace*{10mm}
\includegraphics[height=110mm]{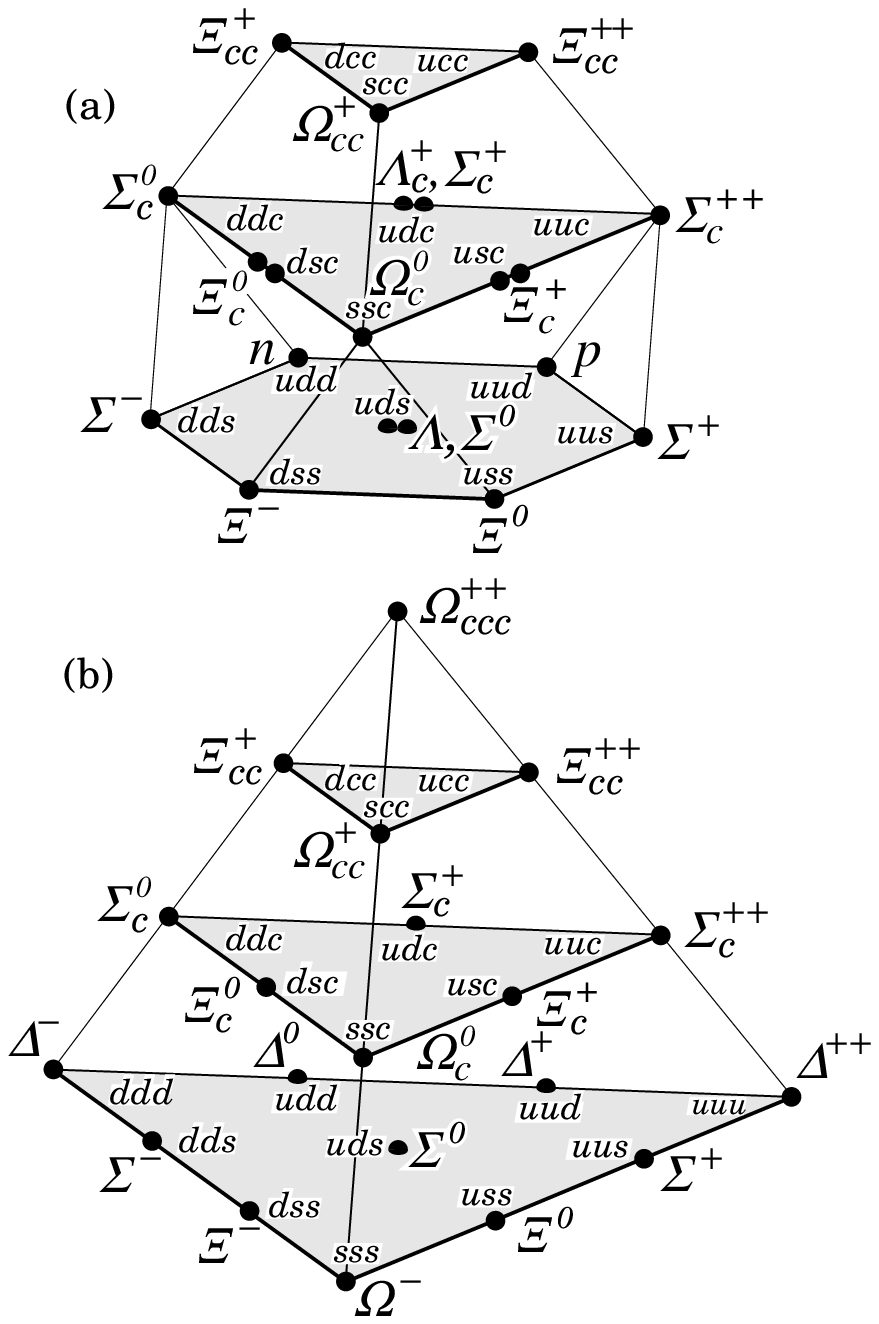}
\caption{\label{fig:charm-hadrons}Diagrams showing the 16-plets for the pseudoscalar and vector mesons on the left, where one can see non-charm mesons on the center planes, and diagrams showing the baryon 20-plets (right) made of $u,d,s$ and $c$ quarks, where the mass and charm content increases from the base upwards. Figure derived from~\cite{Beringer:1900zz}.}
\end{figure}

For accurate description of the final hadron spectrum, it is necessary to implement a detailed list of hadron states and their decay branching ratios. Seemingly negligible assumptions about both can lead to significant differences in the results of such analysis. Frequent testing and cross-checks with the Particle Data Book~\cite{Beringer:1900zz} and other available programs (see, e.g.,~\cite{Petran:2011aa} and references therein) give us confidence, that the hadron spectrum and decay pattern of hadrons consisting of $u,d$ and $s$ quarks are well described in our program. Our hadron mass spectrum involves all **** and *** resonances. This program update introduces charm mesons depicted in Figure~\ref{fig:charm-hadrons} and charm baryons schematically depicted in Figure~\ref{fig:charm-hadrons} together with their higher mass *** and **** resonances.

Particles, that evaporate from a hot boiling quark--gluon `soup' statistically according to the accessible phase space can be described by the SHM. In this scenario, one expects the final hadron state near, but not generally in chemical equilibrium. In the case of more dynamical evolution and sudden hadronization, the final hadron state can be out of chemical equilibrium irrespective of the fireball being or not being chemically equilibrated. Very slow hadronization process, in which all quark flavors have time to (re-)equilibrate in the hadron phase, can also be described statistically, the different scenarios will be reflected by the values of model parameters.

SHARE with CHARM introduces charm statistical hadronization. Charmonium production, an important subtopic, has a long and colorful history. The possibility that the deconfined QGP phase is suppressing the primordial direct charmonium yield was proposed as a signature of QGP formation~\cite{Matsui:1986dk}. In absence of absolute normalized yields  the  experimental study involved consideration of relative abundance as function of centrality, $R_{AA}^{J/\psi}$, which result indicated the expected suppression. Once absolute yield of charmonium became available it was recognized that the absolute yield  could in fact be due to chemical equilibrium statistical hadronization~\cite{Gazdzicki:1999rk}. The discovery that one can describe 'onium production near to equilibrium reintroduced the chemical equilibrium hadronization of charm hadrons~\cite{Andronic:2003zv,Andronic:2008gm} into consideration. 

Finding the charm abundance in chemical equilibrium can be the result of an analysis performed with our program. However, we view the charm yield as arising from a long and complex evolution in QGP. Charm is produced in hard parton scattering very early in the collision~\cite{Schroedter:2001rh}. The yield of charm may evolve from its creation in the initial collision until freeze-out, see Figures 35 and 36 in Ref.\cite{Rafelski:1996hf} where examples of possible evolution of chemical phase space parameter $\gamma_c$ and charm abundance in the QGP are shown. At high matter and charm densities achieved at LHC, charm may be depleted via annihilation considering the long fireball evolution time span, and when temperature is low enough, hadrons emerge produced in coalescence processes~\cite{Schroedter:2000ek,BraunMunzinger:2000px,Thews:2000rj,Gorenstein:2000ck,Kostyuk:2003kt}.

Charm quarks are quite massive, about an order of magnitude above the expected freeze-out temperature. Therefore, charm quarks may have on average smaller velocity of expansion than the light ($u,d,s$) quarks and `fall behind'. As the size of charm particles is smaller, it is natural to assume a higher decoupling temperature $T_{charm}$, a feature we also introduce in this program upgrade and which called for an external CHARM computational module.

SHAREv2 introduced event-by-event fluctuations of particle yields, which further enhanced the model capabilities. They can be used to, e.g., falsify or support the SHM description in case the fluctuations and yields cannot be or are described by the same set of thermal parameters. They may also help decide which statistical ensemble is appropriate, and decouple the correlation of certain thermal model parameters. In the following, we introduce grand-canonical ensemble yields and fluctuations, see Section~\ref{sec:yieldANDfluctuations} and~\ref{sec:fluctuations}. 

\section{Statistical hadronization model in a nutshell}
For correct evaluation of the final hadron state, one has to calculate the:
\begin{itemize}
\setlength{\itemsep}{0mm}
\setlength{\parskip}{0mm}
\item primary particle yields at chemical freeze-out,
\item charm hadron decays followed by
\item decays of resonances.
\end{itemize}

\subsection{Particle yields and fluctuations}
\label{sec:yieldANDfluctuations}
Using the standard textbook approach for grand-canonical ensemble, every hadron of species $i$ with energy $E_i=\sqrt{m_i^2+p_i^2}$ populates the energy states according to Fermi-Dirac or Bose-Einstein distribution function:
\begin{equation}
\label{eq:distribution}
n_i\equiv n_i\left( E_i \right) =\frac{1}{\Upsilon_i^{-1}\exp\left( E_i/T \right)\pm 1},
\end{equation}
where the upper sign corresponds to fermions and the lower one to bosons. The fugacity $\Upsilon_i$ of the $i$-th hadron species is described in detail below in Section~\ref{sec:chemistry}. Then the hadron species $i$ yield will correspond to the integral of the distribution function (Eq.\ref{eq:distribution}) over the phase space multiplied by the hadron spin degeneracy $g_i = (2J_i+1)$ and volume $V$
\begin{equation}
\label{eq:yield}
\langle N_i \rangle \equiv \langle N_i  (m_i,g_i,V,T,\Upsilon_i)\rangle  = g_i V \int\frac{\mathrm{d}^3p}{(2\pi)^3} \, n_i.
\end{equation}

The fluctuation of the yield (Eq.\ref{eq:yield}) can be calculated as:
\begin{equation}
\label{eq:fluctuation}
\left\langle (\Delta N_i)^2 \right\rangle 
= \Upsilon_i \left.\frac{\partial\langle N_i \rangle}{\partial\Upsilon_i}\right|_{T,V} = g_i V
\int \frac{\mathrm{d}^3p}{(2\pi)^3} \, n_i  \left( 1 \mp n_i  \right).
\end{equation}
It is more practical for numerical computation to express the above equations (Eq.\ref{eq:yield},\ref{eq:fluctuation}) as an expansion in modified Bessel functions ($W(x)\equiv x^2 K_2(x)$) as
\begin{align}
\langle N_i \rangle &= \frac{g_i V T^3}{2\pi^2}\sum\limits_{n=1}^\infty\frac{(\pm 1)^{n-1}\Upsilon_i^n}{n^3}{W}\left(\frac{nm_i}{T}\right),\label{eq:yieldexpansion} \\
\left\langle (\Delta N_i)^2 \right\rangle &= \frac{g_i V T^3}{2\pi^2}\sum\limits_{n=1}^\infty\frac{(\pm 1)^{n-1}\Upsilon_i^n}{n^3}\binom{2+n-1}{n}{W}\left(\frac{nm_i}{T}\right)\label{eq:fluctuationexpansion}.
\end{align}
These expansions can be calculated to any desired accuracy as long as the integrals (Eq.\ref{eq:yield},\ref{eq:fluctuation}) converge; for bosons one has to make sure that $\Upsilon_i\exp(-m_i/T)<1$, otherwise the yield integral $\langle N_i \rangle$ diverges. For heavy ($m \gg T$) particles, such as charm hadrons, Boltzmann distribution is a good approximation, i.e., it is sufficient to evaluate the first term of the expansion in Eq.~\ref{eq:yieldexpansion}, which is indeed implemented in the CHARM module to reduce computation time at no observable loss of precision.

To evaluate the yield of hadron resonance with finite width $\Gamma_i$, one has to weigh the yield (Eq.\ref{eq:yield}) by the resonance mass using the Breit-Wigner distribution:
\begin{equation}
\label{eq:yieldwithwidth1}
\langle \tilde{N}_i^\Gamma \rangle = \int \mathrm{d}M\, \langle N_i(M,g_i,T,V,\Upsilon_i) \rangle \frac{1}{2\pi}\dfrac{\Gamma_i}{(M-m_i)^2+\Gamma_i^2/4}\quad\longrightarrow \langle N_i \rangle \text{ for } \Gamma_i\rightarrow 0.
\end{equation}
Using energy independent width implies a finite probability of the resonance being formed with unrealistically small mass. To mitigate this unphysical scenario, one has to use the energy dependent resonance width. The resonance decay energy threshold is a limiting factor in the accessible energy phase space. The partial width of a decay channel $i\to j$ can be well approximated by 
\begin{equation}
\label{eq:partialwidth}
\Gamma_{i \to j}(M)=b_{i\to j}\Gamma_i \left[1-\left(\frac{m_{ij}}{M}\right)^2 \right]^{l_{ij}+1/2}\qquad \text{ for } M>m_{ij},
\end{equation}
where $b_{i\to j}$ is the decay channel branching ratio, $m_{ij}$ is the decay threshold (i.e., sum of the decay product masses) and $l_{ij}$ is the angular momentum released in the decay. The total energy dependent width is then calculated using the partial widths (Eq.~\ref{eq:partialwidth}) for all decay channels of the resonance in question as
\begin{equation}
\Gamma_i(M) = \sum\limits_j\Gamma_{i\to j}(M).
\end{equation}
For a resonance with a finite width, we can then replace Eq.~\ref{eq:yieldwithwidth1} by
\begin{equation}
\label{eq:yieldwithwidth2}
\langle N_i^\Gamma \rangle = \frac{1}{A_i} \sum\limits_j\int\limits_{m_{ij}}^\infty \mathrm{d}M\,\langle N_i(M,g_i,T,V,\Upsilon_i) \rangle \dfrac{\Gamma_{i\to j}(M)}{(M-m_i)^2+\Gamma_i(M)^2/4},
\end{equation}
where $A_i$ is a normalization constant equal to
\begin{equation}
A_i = \sum\limits_j\int\limits_{m_{ij}}^\infty \mathrm{d}M\,\dfrac{\Gamma_{i\to j}(M)}{(M-m_i)^2+\Gamma_i(M)^2/4}.
\end{equation}
Eq.~\ref{eq:yieldwithwidth2} is the form used in the program to evaluate hadron resonance yield whenever calculation with finite width is required. Note, that yield evaluation with finite width is implemented only for hadrons with no charm constituent quark, zero width ($\Gamma_i = 0$) is used for all charm hadrons.

\subsection{Quark chemistry of the hadron state}
\label{sec:chemistry}
The fugacity of hadron states affects yields of different hadrons based on their quark content. It can be calculated from the individual constituent quark fugacities. In the most general case, for a hadron consisting of $N_u^i, N_d^i ,N_s^i$ and $N_c^i$ up, down, strange and charm 
quarks respectively and $N_{\bar{u}}^i,N_{\bar{d}}^i,N_{\bar{s}}^i$ and $N_{\bar{c}}^i$ anti-quarks, the fugacity can be expressed as
\begin{equation}
\label{eq:fugacity}
\Upsilon_i = (\lambda_u\gamma_u)^{N_u^i}(\lambda_d\gamma_d)^{N_d^i}(\lambda_s\gamma_s)^{N_s^i}(\lambda_c\gamma_c)^{N_c^i}
(\lambda_{\bar{u}}\gamma_{\bar{u}})^{N_{\bar{u}}^i}(\lambda_{\bar{d}}\gamma_{\bar{d}})^{N_{\bar{d}}^i}(\lambda_{\bar{s}}\gamma_{\bar{s}})^{N_{\bar{s}}^i}(\lambda_{\bar{c}}\gamma_{\bar{c}})^{N_{\bar{c}}^i},
\end{equation}
where $\gamma_f$ is the phase space occupancy of flavor $f$ and $\lambda_f$ is the fugacity factor of flavor $f$. Note, that we allow for non-integer quark content to account for states like $\eta$ meson, which is implemented as $\eta=0.55(u{\bar{u}}+d{\bar{d}}) + 0.45s{\bar{s}}$ in agreement with~\cite{Li:2007xf}.
It can be shown that for quarks and anti-quarks of the same flavor
\begin{equation}
\gamma_f = \gamma_{\bar{f}}\qquad\text{ and }\qquad \lambda_f = \lambda_{\bar{f}}^{-1},
\end{equation}
which reduces the number of variables necessary to evaluate the fugacity to a half.

It is a common practice to take advantage of the isospin symmetry and treat the two lightest quarks ($q = u,d$) using light quark and isospin phase space occupancy and fugacity factors which are obtained via a transformation of parameters:
\begin{equation}
\gamma_q = \sqrt{\gamma_u\gamma_d},\qquad \gamma_3=\sqrt{\frac{\gamma_u}{\gamma_d}}\label{eq:gammas},
\end{equation}
with straight forward backwards transformation
\begin{equation}
\gamma_u = \gamma_q\gamma_3,\qquad \gamma_d = \gamma_q/\gamma_3,
\end{equation}
and similarly for the fugacity factors
\begin{align}
\lambda_q = \sqrt{\lambda_u\lambda_d},\qquad \lambda_3=\sqrt{\frac{\lambda_u}{\lambda_d}}\label{eq:lambdas},\\
\lambda_u = \lambda_q\lambda_3,\qquad \lambda_d = \lambda_q/\lambda_3.
\end{align}

Chemical potentials are closely related to fugacity, one can express an associated chemical potential $\mu_i$ for each hadron species $i$ via
\begin{equation}
\Upsilon_i = e^{\mu_i/T}.
\end{equation}
It is more common to express chemical potentials related to conserved quantum numbers of the system, such as baryon number $B$,  strangeness $s$, third component of isospin $I_3$ and charm $c$ : 
\begin{align}
\mu_B &= 3T \log \lambda_q, \label{eq:mub}\\
\mu_S &= T \log \lambda_q/\lambda_s,\label{eq:mus}\\
\mu_{I_3} &= T \log \lambda_3, \label{eq:mu3}\\
\mu_C &= T \log \lambda_c\lambda_q,\label{eq:muc}.
\end{align}
(Notice the inverse definition of $\mu_S$, which has historical origin and is a source of frequent mistake).

\subsubsection{Charm chemistry}

While charm hadrons are well described by the above framework along with the other three quark flavors, we follow slightly different approach in determining the charm chemical parameters. First, we consider only symmetric charm+anti-charm pair production (and/or annihilation). At LHC, for which SHARE with CHARM is optimized, $\mu_B$ is very small and therefore the charm chemical potential is vanishing, and charm fugacity factor is effectively unity, $\lambda_c = 1$. This implies that the number of charm quarks and anti-quarks is the same, $N_c = N_{\bar c} = N_{c+\bar{c}}/2$. We determine the charm phase space occupancy $\gamma_c$ following the approach of~\cite{Kuznetsova:2006bh}, where the number of charm quarks $N_c$ is given (as a model parameter) and $\gamma_c$ is found by solving
\begin{align}
\label{eq:gamma_c}
\langle N_{c+\bar{c}} \rangle &= \gamma_c\left(\gamma_q \langle N^{eq}_{qc} \rangle + \gamma_s \langle N^{eq}_{sc} \rangle
+ \gamma_q^2\langle N^{eq}_{qqc} \rangle + \gamma_s\gamma_q\langle N^{eq}_{cqs} \rangle + \gamma_s^2\langle N^{eq}_{ssc} \rangle
\right)\nonumber\\
&+ \gamma_c^2 \left(\langle N^{eq}_{cc} \rangle
+ \gamma_q\langle N^{eq}_{ccq} \rangle
+ \gamma_s\langle N^{eq}_{ccs} \rangle \right)\nonumber\\
&+ \gamma_c^3\langle N^{eq}_{ccc} \rangle,
\end{align}
where $\langle N^{eq}_{ijk} \rangle$ resp. $\langle N^{eq}_{kl} \rangle$ is the sum of equilibrium yields of baryons with quark content $ijk$ and $\bar{i}\bar{j}\bar{k}$, resp. mesons with quark content $k\bar{l}$ and $\bar{k}l$. For instance, $\langle N^{eq}_{cu} \rangle$ includes $D^0$, $\overline{D}^0$, $D^*(2007)^0$, $\overline{D}^*(2007)^0$, etc. (Note that only in the Eq.~\ref{eq:gamma_c}, $\langle N^{eq}_{cc} \rangle$ denotes the sum of charmonium yields, whereas $N_{c\bar{c}}$ everywhere else in the text denotes the number of charm+anti-charm quarks.) Even though Eq.~\ref{eq:gamma_c} is cubic in $\gamma_c$ and has generally three solutions for $\gamma_c$, for physical values of all quantities involved, only one of the solutions is positive and real and is accepted as the value of $\gamma_c$. 

The hadronization of charm itself is a new phenomenon in the physics of heavy-ion collisions and very little is known about this process. Predictions for the amount of charm created in heavy-ion collisions at LHC expect $123\pm77$ charm+anti-charm quark pairs created in a central Pb--Pb collision~\cite{Nelson:2012bc}. We expect this amount to be slightly modified by annihilation and not very abundant thermal production of charm quarks during the expansion of the fireball. Massive charm quarks may expand slower outwards from the primary vertex of the collision. During hadronization, they may find themselves at a point within the fireball at slightly higher temperature. In this case, charm would populate the charm hadron phase space at a temperature above that of the light flavors. We introduce the charm hadronization temperature $T_{charm}$ and use it in Eq.~\ref{eq:yieldexpansion} to calculate charm hadron yields and in Eq.~\ref{eq:gamma_c} when determining the value of $\gamma_c$. The ratio of charm to light hadronization temperature is a newly introduced model parameter, see Section~\ref{sec:parameters}.

\subsection{Resonance decays}
\label{sec:decays}
During the evaluation of hadron yields, the program first calculates the event-by-event average yields and fluctuations at hadronization treating resonances as particles with well defined mass. These quantities are in general different from experimentally observed yields and fluctuations. The resonances decay rapidly after the freeze-out and feed lighter resonances and stable particle yields. The final stable particle yields are obtained by allowing all resonances to decay sequentially from the heaviest to the lightest and thus correctly accounting for resonance cascades. Final yield of each hadron $i$ is then a combination of primary production and feed from resonance decays
\begin{equation}
\label{eq:decayfeed}
\langle N_i \rangle = \langle N_i \rangle_\mathrm{primary} + \sum\limits_{j\neq i}B_{j\to i}\langle N_j \rangle,
\end{equation}
where $B_{j\to i}$ is the probability (branching ratio) that particle $j$ will decay into particle $i$. Applied recursively, Eq.~\ref{eq:decayfeed} reproduces the experimentally observed yields.

For non-charm hadrons, all decay channels with branching ratio $\geq 10^{-2}$ were accepted, but the higher number of charm hadron decays (a few hundred(!) in some cases) with smaller branching ratios required to accept all decay channels with branching ratio $\geq 10^{-4}$. Since charm hadrons in a lot of cases decay into more than three particles, a different approach in implementing them has to be used, see further in Section~\ref{sec:HFfeed}. There is still a lot of uncertainty in charm decay channels. Some of them are experimentally difficult to confirm, but required by, e.g., the isospin symmetry and had to be added by hand for several charm hadrons. For example, a measured $\Lambda_c^+$ decay channel 
\begin{equation}
\label{eq:LambdaCdecay}
\Lambda_c^+ \to p\overline{K}^0\pi^0 \qquad (3.3\pm1.0)\%,
\end{equation}
is complemented by the unobserved isospin symmetric channel 
\begin{equation}
\label{eq:LambdaCdecaySymmetric}
\Lambda_c^+ \to n\overline{K}^0\pi^+ \qquad (3.3\pm1.0)\%,
\end{equation}
with the same branching ratio.

The influence of resonance feed-down on fluctuations is the following:
\begin{equation}
\label{eq:decayfluct}
\langle (\Delta N_{j\to i})^2 \rangle = B_{j\to i}(\mathcal{N}_{j\to i} - B_{j\to i})\langle N_j \rangle \,+\, B_{j\to i}^2\langle(\Delta N_j)^2\rangle.
\end{equation}
The first term corresponds to the fluctuations of the mother particle $j$, which decays into particle $i$ with branching ratio $B_{j\to i}$. $\mathcal{N}_{j\to i}$ is the number of particles $i$ produced in the decay of $i$ (inclusive production) so that $\sum_iB_{j\to i} = \mathcal{N}_{j\to i}$. For nearly all decays of almost all resonances $\mathcal{N}_{j\to i}=1$, however, there are significant exceptions to this including production of multiple $\pi^0$, such as $\eta\to 3\pi^0$. The second term in Eq.~\ref{eq:decayfluct} corresponds to the fluctuation in the yield of the mother particle (resonance).

\subsection{Fluctuations --- volume fluctuations, fluctuations of ratios and finite acceptance}
\label{sec:fluctuations}
In most recent heavy-ion experiments, particle yields and fluctuations are measured in a limited kinematic domain, usually a well defined rapidity range around $y=0$ (central rapidity). Results are then reported per unit of rapidity, e.g., particle yields are $dN_i/dy$. The acceptance domain is in the boost invariant limit equivalent to a configuration space sub-volume~\cite{Torrieri:2005va} and it follows that both particle ratios and fluctuations satisfy:
\begin{equation}
\label{eq:acceptance}
\frac{\langle N_i\rangle}{\langle N_j\rangle}=\frac{dN_i/dy}{dN_j/dy},
\end{equation}
and the scaled variance $\sigma_X^2$ of quantity $X$ defined as
\begin{equation}
\sigma_X^2 = \frac{\langle (\Delta X)^2 \rangle }{\langle X \rangle} = \frac{\langle X^2\rangle - \langle X \rangle^2}{\langle X \rangle},
\end{equation}
will be given by
\begin{equation}
\sigma_{N_i}^2 = \frac{d\sigma_{N_i}^2}{dy}.
\end{equation}

The evaluation of grand canonical ensemble (GCE) fluctuations (Eq.~\ref{eq:fluctuationexpansion}) neglects the fluctuations of volume. These are accounted for by dividing the observed fluctuation into an extensive and intensive part as follows:
\begin{equation}
\langle(\Delta X)^2\rangle \approx \langle (\Delta x)^2\rangle\langle V \rangle^2 + \langle x \rangle^2\langle (\Delta V)^2\rangle,\label{eq:volumefluctuation}
\end{equation}
where $\langle x\rangle$ and $\langle x^2 \rangle$ can be calculated using the above equations in this section.
Volume fluctuations $\langle (\Delta V)^2\rangle$ are difficult to describe in a model independent way and thus the suggested procedure to avoid this problem is to choose observables independent of volume fluctuations. Observables for which $\langle x\rangle^2 \ll \langle(\Delta x)^2\rangle$ are good candidates. Event-by-event fluctuations of particle ratios are even better choice as they are volume fluctuation independent by construction. With a complete decay tree, the fluctuations of particle ratios can be evaluated using the numerator's and denominator's fluctuations. However, one has to keep in mind that resonance decays produce both fluctuations and correlations, as the decays can feed both the numerator and the denominator. For the variance of a ratio of two particles $N_1/N_2$, one should use
\begin{equation}
\label{eq:ratiocorrelation}
\sigma_{N_1/N_2}^2 = \frac{\langle (\Delta N_1)^2\rangle}{\langle N_1 \rangle^2}
                   + \frac{\langle (\Delta N_2)^2\rangle}{\langle N_2 \rangle^2}
                   - 2\frac{\langle \Delta N_1 \Delta N_2\rangle}{\langle N_1 \rangle\langle N_2 \rangle}.
\end{equation}
The last correlation term depends on the resonance decays into both particles as
\begin{equation}
\langle \Delta N_1 \Delta N_2\rangle = \langle N_1N_2 \rangle - \langle N_1\rangle\langle N_2 \rangle
       \simeq \sum\limits_j B_{j\to 1,2}\langle N_j\rangle.
\end{equation}
Even though $\sigma_N^2$ is independent of the average system volume $\langle V\rangle$, the variance of a ratio acquires dependence on it since ratio fluctuations scale with $\langle N \rangle^{-1}$. An analysis incorporating particle fluctuations should therefore include some particle yields data and system volume is strongly suggested as a free fit parameter (see Section~\ref{sec:parameters} for technical details on how to accomplish this).

Most common way to separate detector acceptance effects from physics is to evaluate the fluctuations in `mixed' events, where particles from distinct events are combined. By construction, such fluctuations contain only detector acceptance effects as the particles themselves are not correlated in any other way. Hence, the normalized `static' fluctuation $\sigma_\mathrm{stat}^2$ is determined only by a trivial Poisson contribution and the detector effects. In the SHM, the static fluctuation is 
\begin{equation}
\left(\sigma_{N_i}^2\right)_\mathrm{stat} = 1.
\end{equation}
The correlation term, $\langle \Delta N_1 \Delta N_2\rangle$ of the particle ratio $N_1/N_2$ in mixed events vanishes in Eq.~\ref{eq:ratiocorrelation} and hence the fluctuation of the ratio simplifies to 
\begin{equation}
\left(\sigma_{N_1/N_2}^2\right)_\mathrm{stat} = \frac{1}{\langle N_1 \rangle} + \frac{1}{\langle N_2 \rangle}.
\end{equation}
The dynamical fluctuation $\sigma^2_\mathrm{dyn}$ defined by 
\begin{equation}
\label{eq:fluct-dyn}
\sigma^2_\mathrm{dyn} = \sqrt{\sigma^2 -\sigma_\mathrm{stat}^2},
\end{equation}
corresponds to the difference of the directly measured fluctuation $\sigma^2$ and the static fluctuation from mixed events $\sigma_\mathrm{stat}^2$ and can be shown to be independent of detector acceptance~\cite{Pruneau:2002yf}. This makes the $\sigma^2_\mathrm{dyn}$ more robust to compare with fluctuation estimates from SHM.

Mixed event particles are uncorrelated and hence mixed event techniques cannot account for detector acceptance effects while evaluating particle correlations. The Eqs.~\ref{eq:decayfeed} and~\ref{eq:decayfluct} need to use a corrected branching ratios $B_{j\to i} \rightarrow \alpha_{j\to i}B_{j\to i}$, and consequently, Eq.~\ref{eq:ratiocorrelation} needs to be updated as well to read:
\begin{equation}
\label{eq:ratiocorrelation-modified}
\sigma_{N_1/N_2}^2 = \frac{\langle (\Delta N_1)^2\rangle}{\langle N_1 \rangle^2}
                   + \frac{\langle (\Delta N_2)^2\rangle}{\langle N_2 \rangle^2}
                   - 2\alpha_{12}\frac{\langle \Delta N_1 \Delta N_2\rangle}{\langle N_1 \rangle\langle N_2 \rangle}.
\end{equation}
The first correction factor we introduced, $\alpha_{j\to i}$, correspond to the probability that particle $i$ will end up in the detector acceptance provided that particle $j$ is also inside the region. The second correction, $\alpha_{12}$, corresponds to the probability that both decay products are within the detector acceptance. For boost invariant system with full azimuthal coverage, $\alpha_{j\to i} = 1$, since the particles leaving the detector acceptance will be balanced by those entering it. Unlike $\alpha_{12}$, which in general is $\alpha_{12} < 1$, since for resonances outside the detector acceptance with one of the decay products entering the detector acceptance, the other cannot enter it due to momentum conservation. In practice, this is necessary only at RHIC (much less at LHC) for some weak decays, which are experimentally distinguishable from primary particles. In the program, we offer the option to enter the correction factor $\alpha_{12}$ for any resonance decay as an input parameter, see Section~\ref{sec:weakdecays} for details how to accomplish this.

\section{SHARE with CHARM program structure}
\begin{figure}[!tb]
\includegraphics[width=\columnwidth]{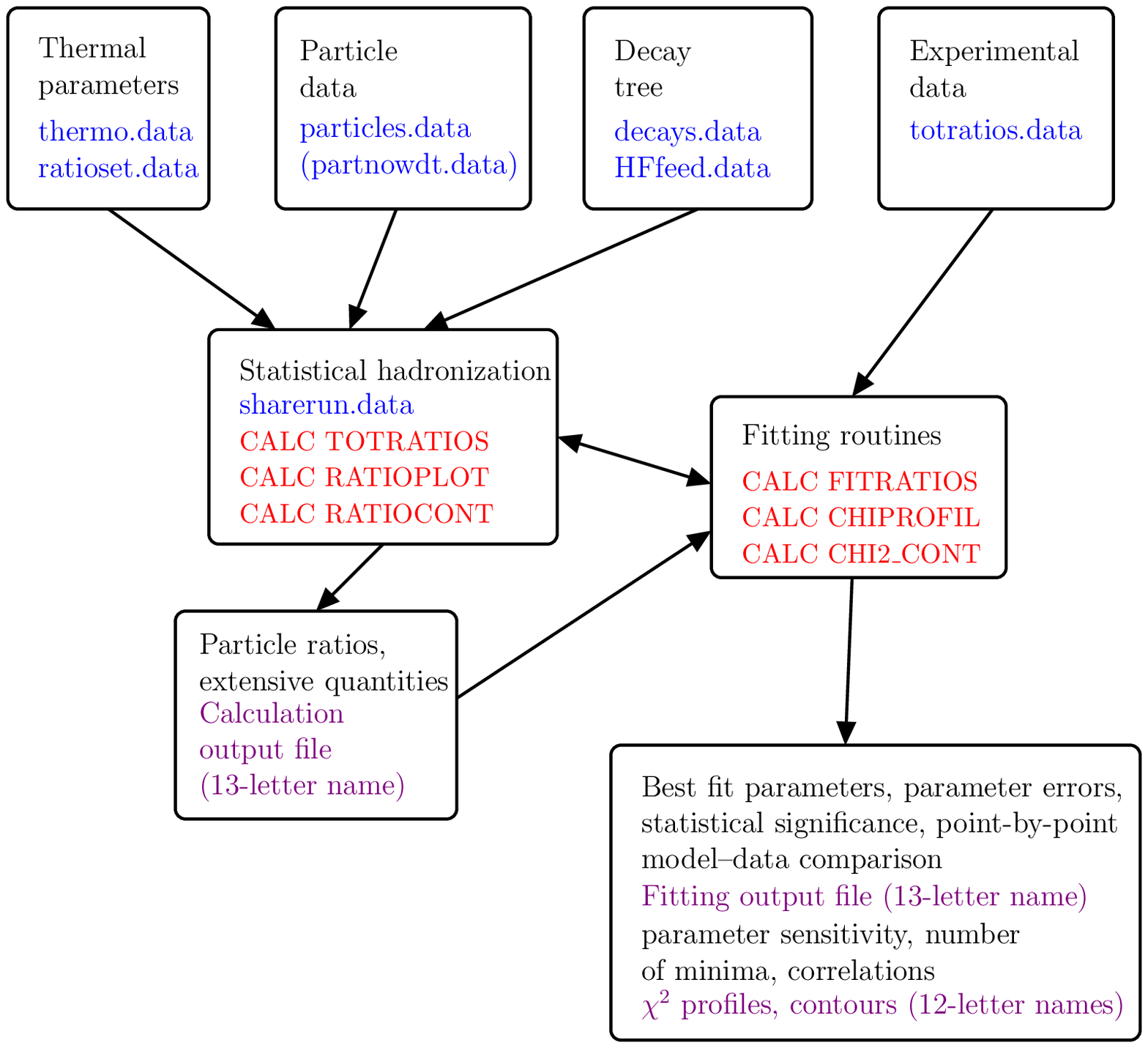}
\label{fig:programstructure}
\caption{(color online) Schematics of the SHARE with CHARM program structure. Default input file names are written in blue, program commands in red and output files in purple.}
\end{figure}

The basic structure of the program is depicted in Figure~\ref{fig:programstructure}. It requires a total of six input files containing list of particles, decay tree, and model parameters. The program can perform a multitude of commands, which are read at run time from the file (\texttt{sharerun.data}). 
The computational and fitting block will perform commands in the run file as entered one after another generally independently of each other. Each command produces an output into a separate file named by the user. The mandatory input files the user has to provide are (default filenames listed):
\begin{itemize}
\setlength{\parskip}{0mm}
\item \texttt{particles.data} (14--letter filename) --- list of particle properties,
\item \texttt{decays.data} (11--letter filename) --- list of non-charm hadron decays,
\item \texttt{HFfeed.data} (constant filename) --- list of inclusive branching ratios of charm hadrons,
\item \texttt{thermo.data} (11--letter filename) --- list of model parameters,
\item \texttt{ratioset.data} (13--letter filename) --- list of model parameter ranges,
\item \texttt{totratios.data} (14--letter filename) --- experimental data and physical properties requested in the output
\item \texttt{sharerun.data} (constant filename) --- the driving file with a list of commands to perform,
\end{itemize}
and an optional input file with weak feed-down corrections
\begin{itemize}
\item \texttt{weak.feed} (9--letter filename) --- list of weak decay feed-down corrections $\alpha_{12}$, as described in Section~\ref{sec:fluctuations}.
\end{itemize}
If the user does not specify any of the mandatory input files explicitly, the program will look for the input file with the default name and will not run correctly if any of the mandatory input files is missing. In all of these files, the user can enter a comment by starting a line with the pound character, \verb+#+. All subsequent characters after \verb+#+ on the line will be ignored. We will address the structure of each input files below in a separate section. It should be pointed out that all input files are read by the program as fixed format input and it is hence crucial to keep their structure including the number of characters allocated for each record (including blank spaces between the records). The only exception is the charm hadron decay file \texttt{HFfeed.data}, which can include any positive number of blank characters (spaces or tabs) in between the records.

\subsection{Thermal parameters}
\label{sec:parameters}
\subsubsection{List of thermal parameters (11-letter filename, default: \texttt{thermo.data})}
The thermal parameter file contains a list of parameters of the model together with their initial value. All parameter names are 4--letter tags. The \texttt{thermo.data} file has to contain the 12 parameters as in the provided program package. We show in Table~\ref{tab:thermodata} the typical contents of a sample file together with the description of each parameter. The units are GeV and fm$^3$, where applicable.

\begin{table}
\caption{\label{tab:thermodata}Structure of the \texttt{thermo.data} file containing thermal parameters and their respective initial values.}
\begin{center}
\begin{tabular}{p{8ex}p{12ex}p{50ex}}
\hline
Parameter &  Initial Value & Parameter decsription\\\hline
\texttt{norm}&  \texttt{2200.} &  absolute normalization in fm$^3$\\
\texttt{temp}&  \texttt{0.139} &  chemical freeze-out temperature $T$ in GeV\\
\texttt{lamq}&  \texttt{1.0055}&  light quark fugacity factor\\
\texttt{lams}&  \texttt{1.0 }  &  strangeness fugacity factor\\
\texttt{gamq}&  \texttt{1.0}   &  light quark phase space occupancy\\
\texttt{gams}&  \texttt{1.0}   &  strangeness phase space occupancy\\
\texttt{lmi3}&  \texttt{1.00}  &  $I_3$ fugacity factor (Eq.~\ref{eq:lambdas})\\
\texttt{accu}&  \texttt{0.001} &  calculations accuracy\\
\texttt{dvol}&  \texttt{0.}    &  statistical pressure ensemble fluctuations (Eq.~\ref{eq:volumefluctuation}) \\
\texttt{gam3}&  \texttt{1.}    &  $I_3$ phase space occupancy (Eq.~\ref{eq:gammas})\\
\texttt{lamc}&  \texttt{1.}    &  charm fugacity factor (\emph{currently fixed to $\lambda_c=1$})\\
\texttt{Ncbc}&  \texttt{0. }   &  number of $c+\bar{c}$ quarks\\
\texttt{tc2t}&  \texttt{1.}    &  ratio of charm to the light quark hadronization temperature $T_{charm}/T$\\\hline
\end{tabular}
\end{center}
\end{table}

The fugacity factors $\lambda_i$ ($i=q,s,I_3$) may be replaced by their respective chemical potentials according to Eqs.~\ref{eq:mub}--\ref{eq:mu3}. The respective tags are \texttt{mu\_b}, \texttt{mu\_s}, and \texttt{mui3}.

The parameter tags in this file are used throughout the program, so it is highly advisable \emph{not} to change the parameter names, as it may result in unpredictable (if any) results. On the other hand, we encourage the user to change the initial parameter values in this file. The format of every line  comprises a 4--letter parameter tag and 4 spaces followed by the initial value of the respective parameter. The relevant Fortran format statement is \texttt{(A4,4X,F31.19)}.

\subsubsection{Fit parameter ranges (13-letter filename, default: \texttt{ratioset.data})}
\label{sec:ratioset}
The \texttt{ratioset.data} file defines the ranges of parameters that will be varied during a fit and which parameters will be kept constant. For each parameter, this file contains the lower and upper limits of its range and the initial step size during minimization with MINUIT. Typical format of this file is shown in Table~\ref{tab:ratioset}. The last column is each row states if the parameter is to be fitted within the specified range (\texttt{1}), or kept constant (\texttt{0}). All parameters defined in the thermal parameters file \texttt{thermo.data} have to be defined in the \texttt{ratioset.data} file as well, however, the parameters do not have to be in the same order in both files. Every line in this file must be compatible with the following format: \texttt{(A4,3X,2F7.1,F7.5,I4)}\\[2mm]
If the upper and lower limits for a parameter are equal and the parameter is to be fitted, MINUIT will consider this parameter unconstrained during a fit. In case the user chooses to fit only experimental particle ratios and densities, the volume is automatically kept constant. We have found that the order of parameters, which is maintained during MINUIT calls, affect the quality (speed, reliability) of the resulting fit. The user is advised to input the most important fit parameters first, usually \texttt{temp} and \texttt{norm}. Second, highly correlated parameters should be placed one right after the other. Note, that some fit parameters (e.g., strangeness fugacity factor $\lambda_s$) can be a result of a conservation law (e.g., of strangeness). For details about possible conservation law implementations, see Section~\ref{sec:conservationlaws}.

\begin{table}
\caption{\label{tab:ratioset}Structure of the \texttt{ratioset.data} file.}
\ttfamily\selectfont
\begin{center}
\begin{tabular}{*{5}{l}}\hline
\textnormal{Tag} & \textnormal{Lower limit} & \textnormal{Upper limit} & \textnormal{Initial step} & \textnormal{Fit?=0/1}\\\hline
norm  &  9.  &  9999. & 1.0  &    1\\
temp  &  0.1 &   0.3  &  0.1 &    1\\
lamq  &  0.1 &   90.  &  0.1 &    1\\
lams  &  0.9 &   90.  &  0.1 &    1\\
gamq  &  0.2 &   2.5  &  0.1 &    1\\
gams  &  0.1 &    9.  &  0.1 &    1\\
mui3  &  0.01&   92.5 &  0.1 &    1\\
Ncbc  &  0.  &   400. &  1.0 &    0\\
dvol  &  0.  &   10.  &  0.1 &    0\\
gam3  &  0.5 &   1.5  &  0.1 &    0\\
tc2t  &  0.5 &   4.   &  0.1 &    0\\
lamc  &  0.9 &   900. &  0.1 &    0\\
\hline
\end{tabular}
\end{center}
\end{table}

\subsection{Particle properties data file (14--letter filename, default	: \texttt{particles.data})}
\label{sec:particlelist}
This input file contains the list of particles and its properties, specifically the particle name, mass, width, spin, isospin, quark content and Monte Carlo (MC) code where available. The structure of every line in the file is the following:\\[2mm]
\texttt{name\hfil  mass\hfil  width\hfil  spin\hfil  I\hfil  I$_3$\hfil  u\hfil  d\hfil  s\hfil  au\hfil  ad\hfil  as\hfil  c\hfil  ac\hfil  b\hfil  ab\hfil  MC\\[1mm]}
where\\[1mm]
\begin{tabularx}{\textwidth}{lX}
\texttt{name} & a nine character string identifying the particle,\\
\texttt{mass} & particle mass in GeV/c$^2$,\\
\texttt{width} & particle width in GeV,\\
\texttt{spin} & spin of the particle,\\
\texttt{I} & particle isospin,\\
\texttt{I$_3$} & the third component of the particle isospin,\\
\texttt{u,d,s,c,b} & number of particle constituent quarks, $u,d,s,c$ and $b$ respectively,\\
\texttt{au,ad,as,ac,ab} & number of particle constituent anti-quarks, $\bar{u},\bar{d},\bar{s},\bar{c}$ and $\bar{b}$ respectively,\\
\texttt{MC} & particle MC identifier, where available. Note that not all hadron resonances have a MC code assigned in the standard scheme~\cite{Beringer:1900zz}.\\[2mm]
\end{tabularx}

The particle \texttt{name} defined in the particle list is used as a unique particle identifier throughout the program and we strongly advice not to change the names of particles already provided in the input file that comes as a part of the program package. The following naming convention was chosen and is used for most particles; first two characters of the (9--letter) particle name is an abbreviation of the particle letter followed by 4 numbers representing particle approximate mass (in $\mathrm{MeV/}c^2$) and the last three letters reflect the particle electric charge, e.g., $\Lambda$ can be found in the program as \texttt{Lm1115zer}. The following 3--letter endings are being used:\\[2mm]
\begin{tabularx}{\textwidth}{llX}
\texttt{zer} & `zero'     & neutral particle, charge $0$,\\
\texttt{zrb} & `zero bar' & neutral anti-particle, charge $0$ (neutral anti-baryons,\\
\texttt{plu} & `plus'     & positive particle, charge $+$,\\
\texttt{plb} & `plus bar' & positive anti-particle, charge $-$ (anti-baryons, e.g., the antiproton $\bar{p}^-$),\\
\texttt{min} & `minus'    & negative particle, charge $-$,\\
\texttt{mnb} & `minus bar'& negative anti-particle, charge $+$ (e.g., $\overline{\Omega}^+$), \\
\texttt{plp} & `plus plus'& doubly positive particle, charge $++$ (e.g., $\Delta^{++}$),\\
\texttt{ppb} & `plus plus bar'& doubly positive anti-particle, charge $--$,\\
\texttt{nuc} & `nucleus'  & (hyper-)nucleus with charge above the range of previous endings\\
\texttt{anc} & `anti-nucleus'  & (hyper-)anti-nucleus with charge above the range of previous endings\\[2mm]
\end{tabularx}

When editing the list of particles, the user is strongly encouraged to create a copy of a line in the particle properties file and edit the copied line replacing characters one-to-one to prevent formatting changes. The general expected format of a line is defined by the Fortran 77 format statement:\\
\texttt{(A9,2(4X,F10.7),2X,3(F5.1,4X),10(F6.3,2X),I9)}\\
where `A' is a letter, `X' is a space, `F' is a real number and `I' is an integer number. For more details about format specification as defined by the Fortran 77 standard, see, e.g., Ref.~\citep[section 13]{F77standard}. 

For example, motivated by our recent analysis of LHC data~\cite{Petran:2013dea}, we include a ** resonance $\Sigma(1560)$ in the particle list and decay tree as \texttt{Sg1560???}, but, as a 2-star resonance, we leave it commented out in both. In other words, this resonance is not included in the calculations by default. If the user chooses to include this resonance in the analysis, it is enough to remove the \texttt{\#} character in front of the following lines in the particle list\\
\texttt{\#Sg1560min     1.5530000 \ldots $\qquad\;\longrightarrow\qquad$  Sg1560min     1.5530000 \ldots\\
\#Sg1560mnb     1.5530000 \ldots $\qquad\;\longrightarrow\qquad$  Sg1560mnb     1.5530000 \ldots\\
\ldots} etc.\\
and same in the decay tree file to enable the $\Sigma(1560)\to \Lambda\pi$ decay, see Section~\ref{sec:decayfile-light} below for details on the decay tree structure.

The $\Delta(1232)^{0}$, for instance, will appear in the list as:\\[1mm]
\texttt{
Dl1232zer\hfil 1.232\hfil      0.12\hfil 1.5\hfil 1.5\hfil -0.5\hfil 1.\hfil 2.\hfil 0.\hfil 0.\hfil 0.\hfil 0.\hfil 0.\hfil 0.\hfil 0.\hfil 0.\hfil 2114\\[1mm]}
We allow non-integer quark numbers to accommodate strong interaction flavor mixing, such as the $\eta$ meson mentioned in Section~\ref{sec:chemistry}. Since SHAREv2~\cite{Torrieri:2006xi}, we enlarged the particle list to include all at least *** charm hadrons and resonances and updated those present and increased the total number of particles a user can specify to 1000.  No bottom hadron is at the moment present in the particle list provided with this SHARE program release. The particle input file has been re-designed with future upgrade in mind, that would include calculation of bottom hadrons. Up to this day, no significant amount of bottom hadrons is expected in heavy-ion experiments and hence the list provided with the program does not contain any bottom hadrons.

SHARE with CHARM calculations are relevant for strongly interacting system, where $K^0$ and $\overline{K}^0$ are the relevant states. Electroweak mixing $K^0-\overline{K}^0$ occurs on a longer timescale and hence should be performed at the end of the calculation. For the convenience of the user, we implemented the following name ending shortcuts representing basic algebraic combinations of particles from the list.\\[2mm]
\begin{tabular}{lll}
\texttt{sht} & `short'    & \texttt{sht=$^1/_2$(zer+zrb)}, used to calculate mixed states,\\
             &            &  e.g., the often measured $K_S=\frac{1}{\sqrt{2}}(K^0-\overline{K}^0)$,\\
\texttt{lng} & `long'     & \texttt{lng=$^1/_2$(zer+zrb)}, used to calculate mixed states,\\
             &            &  e.g., the less often measured $K_L=\frac{1}{\sqrt{2}}(K^0+\overline{K}^0)$,\\
\texttt{tot} & `total'    & \texttt{tot=(zer+zrb)}, e.g., \texttt{Lm1115tot} for $\Lambda+\overline{\Lambda}$,\\
\texttt{plt} & `plus total'& \texttt{plt=(plu+plb)}, e.g., \texttt{pr0938plt} for $p+\overline{p}$,\\
\texttt{pbn} & `plus bar net'  & \texttt{pbn=(plu-plb)}, e.g., \texttt{pr0938pbn} for $p-\overline{p}$,\\
\texttt{mnt} & `minus total'& \texttt{mnt=(min+mnb)}, e.g., \texttt{UM1672mnt} for $\Omega+\overline{\Omega}$,\\
\texttt{mnn} & `minus net'  & \texttt{mnn=(min-mnb)}, e.g., \texttt{UM1672mnn} for $\Omega-\overline{\Omega}$,\\
\texttt{plm} & `plus minus'     & \texttt{plm=(plu+min)}, e.g., \texttt{pi0139plm} for $\pi^++\pi^-$,\\
\texttt{pln} & `plus minus net' & \texttt{pln=(plu-min)}, e.g., \texttt{pi0139pln} for $\pi^+-\pi^-$,\\
\texttt{pmb} & `plus minus bar' & \texttt{pmb=(plb+mnb)}, e.g., \texttt{Sg1189pmb} for $\overline{\Sigma}^+ +\overline{\Sigma}^-$.\\[2mm]
\end{tabular}\\
Note, that these combinations are \emph{not} in the particle list as separate entries, they are evaluated during the program run from the particle yields as one of the last steps of the calculation and they are printed in the output if so requested by the user (see Section~\ref{sec:experimentaldata}).

\subsection{Particle decays}
Light hadron decays can be divided into two categories, light hadron ($u,d,s$) 2-body and 3-body decays, and charm hadron decays.
\subsubsection{Light hadron decay tree (11--letter filename, default: \texttt{decays.data})}
\label{sec:decayfile-light}
The treatment of light hadron decays did not change since SHAREv1~\citep{Torrieri:2004zz}. Only 2 and 3-body decays are considered. Every line in \texttt{decays.data} contains one decay channel in the following format:\\[2mm]
\texttt{
Parent  daughter1 daughter2 BR C-G?\\[2mm]
}
for 2-body decays, and\\[2mm]
\texttt{
Parent  daughter1 daughter2 daughter3 BR C-G?\\[2mm]
}
for 3-body decays, where \texttt{Parent} refers to the decaying particle producing daughter particles \texttt{daughterN} with branching ratio \texttt{BR}. The \texttt{C-G?} flag signalizes if the branching ratio should be corrected by a Clebsch-Gordan coefficient (0 = no, 1 = yes). For example, the decay $K^{0*}\to K\pi$ reported to have a branching ratio of $\sim 1.00$ in~\citep{Beringer:1900zz} appears in the decay tree file as two entries:
\begin{verbatim}
Ka0892zer   pi0139min   Ka0492plu   1.0    1
Ka0892zer   pi0135zer   Ka0492zer   1.0    1
\end{verbatim}
whereas other decays, such as decays $\phi\to K^+K^-$ and $\phi\to \pi^+\pi^-\pi^0$, do not need the Clebsh-Gordan coefficients correction, as each combination of daughter particles is reported separately;
\begin{verbatim}
ph1020zer   Ka0492plu   Ka0492min   0.49    0
ph1020zer   pi0139plu   pi0139min   pi0135zer   0.08     0
\end{verbatim}
The general format of 2 and 3-body decays is\\
\texttt{(A9,3X,A9,3X,A9,F9.4,I4)}\\
and\\
\texttt{(A9,3X,A9,3X,A9,3X,A9,F9.4,I4)}\\
respectively. As a general rule, we again suggest that any modifications to this input file are made on one-to-one character replacement basis to prevent formatting discrepancies. Currently, the number of decay channels for each parent particle has been limited to 50. 

\subsubsection{Weak decay corrections (9--letter filename, e.g., \texttt{weak.feed})}
\label{sec:weakdecays}
As discussed in Section~\ref{sec:fluctuations}, feed-down is an important feature of the experimental data to be accounted for within the statistical hadronization model. Limited acceptance of detectors requires experimental correction of the feed-down coefficients reflecting the probability of a resonance being produced within the detector acceptance and the decay product also being in within the detector acceptance region. 

Weak decays, in particular, are susceptible to detector acceptance and efficiency corrections, as they happen at a macroscopic distance from the primary vertex. Weak decay corrections hence comprise a geometrical as well as momentum component. A significant portion of the final particle yield is oftentimes mostly feed-down, e.g., protons are mostly products of hyperon decay, such as $\Lambda\to p\pi$. At RHIC, careful evaluation of the correction factors for each weak decay product was a crucial element to fit the experimental data, whereas at LHC, thanks to a more precise tracking, virtually all weak decays are subtracted from the final hadron yields. 

Decays defined in \texttt{decays.data} file, which violate conservation of strangeness and/or isospin are identified by the program as weak. The weak feed-down correction file can contain any number of weak decays defined in the decay tree (from 0 up to all the weak decays). Consequently, no strong (or electromagnetic) decay corrections are allowed in this file, they are caught as input errors by the program, which exits after reporting the particular decay on screen. Each line in the weak decay file has the following structure for 2-body and 3-body weak decays respectively:\\[2mm]
\resizebox{\columnwidth}{!}{
\begin{minipage}[h]{6.9in}
\texttt{Parent  daughter1  daughter2 all/1st/2nd/cor  coefficient\\
Parent  daughter1  daughter2 daughter3 all/1st/2nd/3rd/cor  coeffi-\\cient
}\\[-1mm]%
\end{minipage}}
\\formatted respectively:\\
\texttt{(A9,3X,A9,3X,A9,3X,A3,F9.4)\\
(A9,3X,A9,3X,A9,3X,A9,3X,A3,3X,F9.4)}\\[2mm]
where \texttt{Parent} and \texttt{daughterN} are the names of particles involved in the decay, similarly to the decay tree file \texttt{decays.data} structure. The decay products do not have to be in the same order as in the decay tree, the program handles any permutation of the decay products (daughter particles). The following 3--letter tag specifies which of the decay products is being corrected. For example, the weak decay $\Lambda\to p\pi^-$ can appear in the weak decay file multiple times specifying a different correction for each of the daughter particles:
\begin{verbatim}
Lm1115zer   pr0938plu   pi0139min   1st   1.0000
Lm1115zer   pr0938plu   pi0139min   2nd   0.0000
\end{verbatim}
The STAR experiment at RHIC filters out the pion, but accepts the proton, as is seen in the above example. The user can specify, whether \texttt{all} daughters will have the same correction \texttt{coefficient}, or only one of them, referring to the \texttt{1st/2nd/3rd} daughter on each line. The last 3--letter tag \texttt{cor} refers to the fractional contribution of the acceptance to the two particle correlation $\langle$\texttt{daughter1 daughter2}$\rangle$ induced by a common parent resonance decay denoted by $\alpha_{12}$ in Eq.~\ref{eq:ratiocorrelation-modified} in Section~\ref{sec:fluctuations}.
SHARE will then renormalize the branching ratio of \texttt{Parent$\to$all/1st/2nd/3rd/cor} by a \texttt{coefficient} when calculating all data points after a given \texttt{weakdecay} statement (see below).

The program may be run without any weak decay feed-down file, in which case the user has to choose from two cases:
\begin{enumerate}
\item All weak decays are accepted in the program, that is all particle yields contain contributions from weak decays. This scenario means, that the weak decays have not been corrected for in the experiment, hence the option is labeled \texttt{UNCORRECT}.
\item No weak decays are accepted, all weak decays are subtracted from the final experimental yields. This option is labeled \texttt{NOWK\_FEED}. This is the default option, unless specified otherwise by the user.
\end{enumerate}
This upgrade of SHARE is tailored to fit data from LHC measured by the ALICE experiment, which publishes results fully corrected for weak decays. The default treatment of weak decays is thus changed from SHAREv2 to `\emph{not} to accept' weak decay feed-down to any particle, the option \texttt{NOWK\_FEED} described above. 

All weak decay file options and user file names should be entered into the experimental data file (\texttt{totratios.data}) with the keyword \texttt{weakdecay} as \texttt{name1} (see Section~\ref{sec:experimentaldata}) followed by one of the options above or user created weak decay feed-down correction file name as follows:
\begin{verbatim}
weakdecay  weak.feed
...
weakdecay  NOWK_FEED
...
weakdecay  UNCORRECT
\end{verbatim}
Experimental data files are read in sequence and hence every weak decay feed-down pattern applies to experimental data points below until another \texttt{weakdecay} statement is encountered. This allows the user to fit data points from multiple experiments with different weak decay feed-down corrections. For detailed information on the general structure of the experimental data file, see Section~\ref{sec:experimentaldata}.

\subsubsection{Charm decay tree (fixed filename \texttt{HFfeed.data})}
\label{sec:HFfeed}
Although charm hadrons decay weakly, life times of charm hadrons are very short due to the charm quark large mass. Therefore, all charm decays are very rapid and are treated separately from the weak decay methods described in the previous section (Section~\ref{sec:weakdecays}). The weakly decaying charm hadrons can have up to several hundred decay channels with similar branching ratios and up to six decay products. Without any dominant decay channel present, we have to include all these concurrent decays. The treatment of light hadron decays described above was thus found unsuitable for the charm region.

In order to save computing time, inclusive branching ratios of charm hadron decays are specified in the charm decay tree. The yields of charm decay products still follows the logic of Eq.~\ref{eq:decayfeed}, however, the sum has significantly less terms because every daughter particle appears only once. Each line of the charm decay file specifies the \texttt{parent} particle followed by one \texttt{daughter} particle and the respective inclusive branching ratio \texttt{BR} in the following format:\\[2mm]
\texttt{Parent daughter BR}\\[2mm]
This file is the only exception among input files that can contain any amount of spaces between the columns.

On an example of $D^{*0}(2007)$, we show the organization of charm decays. The $D^{*0}(2007)$ has two decay channels with branching ratios as follows~\citep{Beringer:1900zz}:
\begin{align}
D^{*0} & \to D^0\pi^0,		& \Gamma_1= 0.619\pm0.029,\nonumber\\
D^{*0} & \to D^0\gamma,	&     \Gamma_2= 0.381\pm0.029,\nonumber
\end{align}
which corresponds to the following in the \texttt{HFfeed.data} file:
\begin{verbatim}
Dc2007zer     Dc1800zer     1
Dc2007zer     pi0135zer     0.619
Dc2007zer     gam000zer     0.381
\end{verbatim}
Every daughter particle type is on a separate line with its respective inclusive branching ratio. One can look at each line in the charm decay file from an equivalent perspective; how many daughter particles of a given type are produced on average after the decay of one parent particle.

\subsection{(Experimental) Values to be calculated (14--letter filename, default: \texttt{totratios.data})}
\label{sec:experimentaldata}
\begin{figure}
\begin{center}
\caption{\label{fig:experimentaldata}A typical structure of a \texttt{totratios.data} file.}
\includegraphics[scale=0.87]{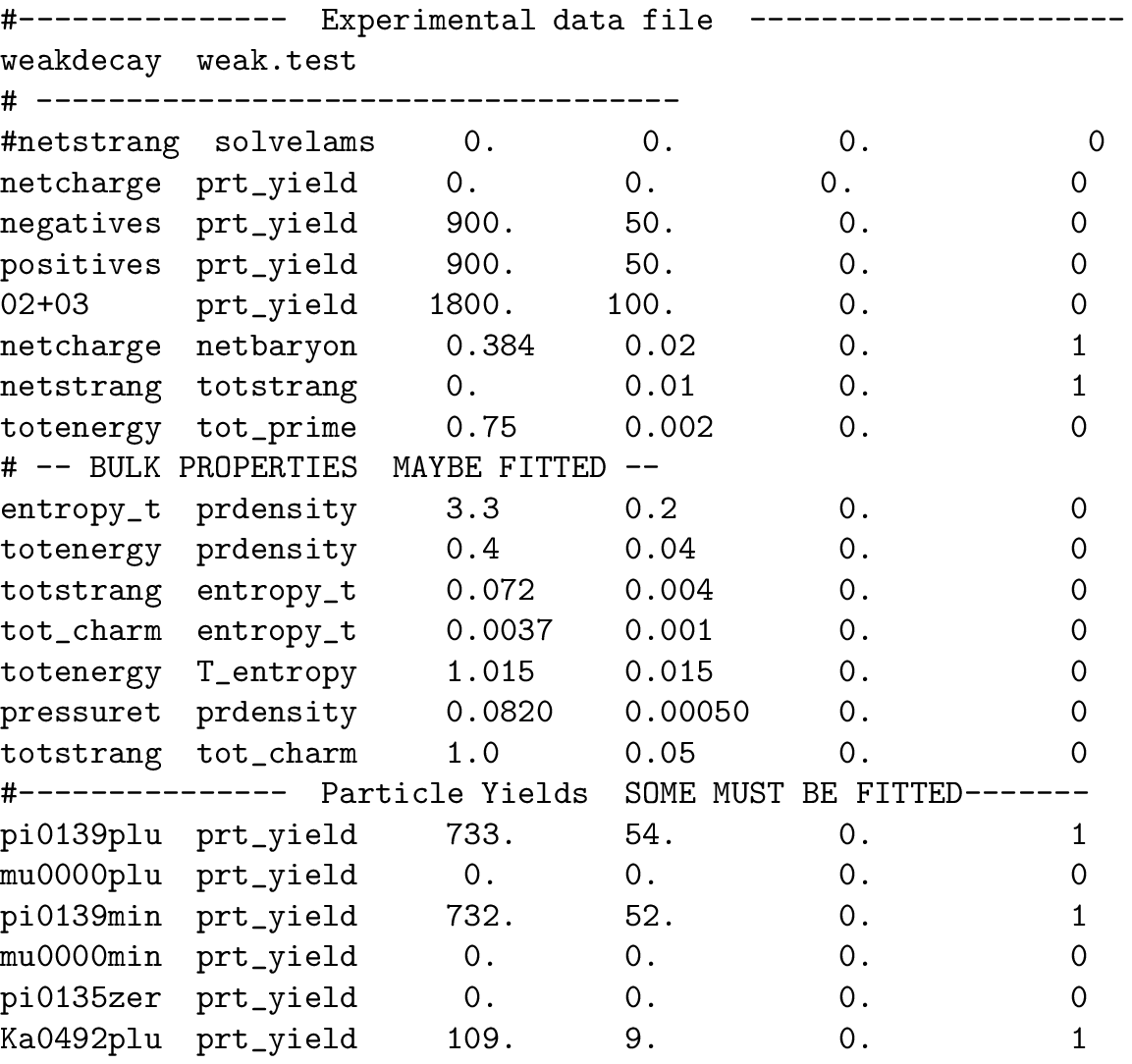}
\end{center}
\end{figure}

Experimental data values and other values of interest to be calculated are defined in the file \texttt{totratios.data}. We provide an example of a typical experimental values file structure in Figure~\ref{fig:experimentaldata}. The general format of every line is \texttt{(A9,2X,A9,3F12.7,I17)}. We would like to point out a common source of error which arises when the decimal point is omitted while entering experimental values in this file. Unfortunately, this is an intrinsic feature of Fortran 77 and cannot be easily mitigated.

\subsubsection{Particle ratios}
\label{sec:particleratios}
Every line in the experimental data file should have the following format:\\[2mm]
\texttt{name1\hfil name2\hfil data\hfil statistical\hfil systematic\hfil fit?(-1/0/1/2)}\\[2mm]
where\\
\begin{tabularx}{\textwidth}{lX}
\texttt{name1}& The first particle (numerator of a ratio), 9--letter name as defined in the particle list (Section~\ref{sec:particlelist}).\\
\texttt{name2}& The second particle (denominator of a ratio), or a tag indicating that the yield or density of \texttt{name1} is entered, 9--letter name.\\
\texttt{data}& The experimental value of this data point.\\
\texttt{statistical}& The statistical error of the data point.\\
\texttt{systematic}& The systematic error of the data point.\\
\texttt{fit?}& This flag specifies if this data point contributes to the evaluation of $\chi^2/\mathrm{ndf}$ of the fit, when set to a positive value (1 or 2). When set to 0, the ratio is not fitted, but calculated and output to the graph file (see Section~\ref{sec:fitting} for details). Values of fit?=-1 or 2 mean that the data point will \emph{not} be output to the graph file. Note, that under typical circumstances, the options 1 and 0 will be sufficient for typical program use.\\
\end{tabularx}

\subsubsection{Particle yields and bulk source properties}
Apart from particle ratios, the user have also the following options of quantities to be fitted, calculated, and printed out in the output.\\[2mm]
{\bf\texttt{name2}} can contain the following 9--letter tags:\\
\begin{tabularx}{\textwidth}{lX}
\texttt{prt\_yield} &	The yield of the first particle, or collective extensive quantity\\
\texttt{prdensity}  &	The density of the first particle or quantity.\\
\texttt{solveXXXX}  &	Evaluation of the parameter \texttt{XXXX} based on first particle or quantity, see Section~\ref{sec:conservationlaws} for details.\\
\texttt{fluctXXXX}   &  Grand-canonical fluctuation, \texttt{XXXX} can contain one of the following: \texttt{\_dyn}, \texttt{dynv}, \texttt{dnch}, \texttt{sgsm}. \\
	& \texttt{fluct\_dyn} to calculate $\sigma_{dyn} = \sqrt{\sigma_{tot}^2 - \sigma_{stat}^2}$ (see Eq.~\ref{eq:fluct-dyn}).\\
	& \texttt{fluctdynv} to calculate $\sigma_{dyn} = \langle N_2 \rangle\, \sigma_{N_1/N_2}$ (see Eq.~\ref{eq:ratiocorrelation-modified}).\\
	& \texttt{fluctdnch} to calculate $\sigma_{dyn} = \langle N_{ch} \rangle\, \sigma_{N_1/N_2}$, the fluctuation scaled by the average number of charged particles $\langle N_{ch} \rangle$. \\
	& \texttt{fluctsgsm} to calculate $\sigma_{dyn} = \langle N_1+N_2 \rangle\, \sigma_{N_1/N_2}$ \\
	& (Other fluctuation options have been discontinued due to their sensitivity to acceptance effects.)\\
\end{tabularx}
\vskip2mm
\noindent{\bf \texttt{name1}} can contain the following 9--letter tags:\\[2mm]
\begin{tabularx}{\textwidth}{lX}
\texttt{negatives}  &  negative particles (weak decay corrections taken into account)\\
\texttt{positives}  &  positive particles (weak decay corrections taken into account)\\
\texttt{chargmult}  &  charged particles (weak decay corrections taken into account)\\
\texttt{neut\_mult}  & neutral particles (weak decay corrections taken into account)\\
\texttt{tot\_multi}  &  total hadron multiplicity (weak decay corrections taken into account)\\
\texttt{tot\_prime}  &  total primary particles multiplicity (no decays considered)\\
\texttt{totstrong}  &  total multiplicity including contributions from strong decays\\
\texttt{totstrang}  &  total strangeness $\langle s + \bar{s}\rangle$\\
\texttt{netstrang}  &  net strangeness $\langle s - \bar{s}\rangle$\\
\texttt{tot\_light}  & total number of light quarks (weak decay corrections taken into account) \\
\texttt{totcharge}  &  total charge $\langle Q + \overline{Q} \rangle$\\
\texttt{netcharge}  &  net charge $\langle Q - \overline{Q} \rangle$\\ 
\texttt{totbaryon}  &  total baryon number, i.e., sum of all baryons and anti-baryons, $\langle B + \overline{B}\rangle$\\
\texttt{netbaryon}  &  net baryon number, i.e., baryons minus anti-baryons $\langle B - \overline{B}\rangle$\\
\texttt{tot\_charm}  &  total charm $\langle c + \bar{c} \rangle$ (corresponding to \texttt{Ncbc}, see Section~\ref{sec:parameters})   \\
\texttt{totenergy}  &  total energy (in GeV)\\[2mm]
\end{tabularx}

The above quantities are calculated \emph{before} the weak decays have occurred. They do not take weak decay acceptances into account, which may produce an apparent violation of strangeness, charge or baryon number. To get values of the above quantities \emph{after} weak decays, one should move the \texttt{net} and \texttt{tot} prefixes to the end of the respective quantity name tags. The equivalents of the above quantities \emph{after} weak decays have occurred are: \texttt{chargetot, chargenet, baryonnet, baryontot, stragetot, strangnet, light\_tot}. These are complemented by the following name tags:\\[2mm]
\begin{tabularx}{\textwidth}{lX}
\texttt{entropy\_t}  & total entropy $S$  \\
\texttt{pressuret}  &  total pressure (in units of energy density, GeV/fm$^3$)\\[2mm]
\end{tabularx}
and the auxiliary\\[2mm]
\begin{tabularx}{\textwidth}{lX}
\texttt{T\_entropy} & $TS$, temperature times entropy\\
\texttt{T\_\_\_\_\_\_\_4} & $T^4/(\hbar c)^3$ (useful to evaluate, for example, the trace anomaly $(\varepsilon-3P)/T^4$)\\
\end{tabularx}\\

Weak decay feed-down correction file or option is indicated in the experimental file by the keyword \texttt{weakdecay}. When this keyword is encountered during the program run as \texttt{name1}, then \texttt{name2} must be either of \texttt{NOWK\_FEED, UNCORRECT}, otherwise \texttt{name2} is treated as the weak decay feed-down filename. The weak decay scheme thus specified applies to all subsequent data points until new \texttt{weakdecay} statement or end of experimental data file.

\subsubsection{Referring to data points and their combinations}
We offer even more flexible handling of data points, we have implemented simple algebraic combinations of data points within a file. The user can refer to a data point defined in the same file by its two-digit line number (not counting comment lines starting with \texttt{\#}). The following snippet of an input file shows how to refer to another data point; line number 8 refers to the $\pi^-$ defined on line 7, however, calculates its density. We also show a more advanced example of how to define and fit the ratio of $(p+\bar{p})/(\pi^-+\pi^+)$. Note that line numbers are not part of the experimental data file, they are provided below for easier orientation in the example.\\
\ttfamily\selectfont
\begin{tabular}{lllllll}
\normalfont{line \#} & \normalfont{input file}\\
\vdots& \vdots\\
\normalfont{7}  & pi0139min & prt\_yield & data & stat & syst & fit? \\
\normalfont{8}  & 07        & prdensity  & data & stat & syst & fit? \\
\normalfont{9}  & pi0139plu & prt\_yield & data & stat & syst & fit? \\
\normalfont{10} & pr0938plu & prt\_yield & data & stat & syst & fit? \\
\normalfont{11} & pr0938plb & prt\_yield & data & stat & syst & fit? \\
   & \multicolumn{6}{l}{\# this is a comment, line number does not increase.}\\
\normalfont{12}  & 10+11     & 07+09 & data & stat & syst & fit? \\[2mm]
\end{tabular}
\normalfont\selectfont

The possible operations are addition, subtraction, multiplication and division accomplished by using \texttt{+,-,X,/} symbols respectively in between two two-digit line numbers as seen in the above example. Note the necessary trailing 0 for single digit line numbers. These operations are limited to two-digit line numbers, although the input file can in general have up to 200 entries (not counting commented lines). Note that this feature uses implicitly recursive code, which may be compiler dependent. It has been found to work as intended on several platforms and compilers we tested SHARE with CHARM on. We provide a sample data file with the program package.

\subsubsection{Statistical and systematic errors}
The systematic and statistical errors can be entered as two separate quantities in SHARE as seen in the above example. This option is created since systematic error is not a random variable that the statistical error is. Systematic error originates from the experimental setup and/or data analysis methods and it can be common to several experimental results, e.g., the efficiency to track strange antibaryons and baryons has same systematic uncertainty, or there could be a strong (anti)correlation, e.g., if an observed particle track is a proton or $K^+$ is not always possible to decide, but if it is one, it cannot be the other. By allowing for systematic and statistical error entry as matter of principle, we prepare for a more complete future treatment of the systematic error including an error correlation analysis.

Once the systematic error correlation matrix function  is known, one must discover combinations of the data which suffer least from the systematic error. The fit than involves a data set in which some of the fitted quantities have a much reduced  systematic error.  At present, the systematic error correlation matrix  is not made available by the experimental groups.  Therefore,  such more detailed error treatment in the fitting procedure is not included in this release of the SHARE with CHARM program suite as a procedure could neither be properly set-up nor tested.

In the current release of the program, if both errors are made available and have been entered separately, they will be added to obtain the total error \texttt{error} of the data point: \texttt{error = stat + syst}.

We note, that as outcome of this procedure, we often see in the study of the RHIC and LHC data that the overall normalization factor $dV/dy$ shows a large and apparently common error of all data, suggesting that all results we interpreted had as input a common systematic tracking efficiency error.

\subsubsection{Conservation Laws}
\label{sec:conservationlaws}
SHARE with CHARM allows the user to solve for a thermal parameter based on a fixed (or experimental) value rather than fit it. The most common application of this feature would be exact conservation of strangeness, $\langle s \rangle = \langle \bar{s} \rangle$, which means numerically solving for $\lambda_s$. In this case, $\lambda_s$ is not a fit parameter anymore, but rather an analytical function of the other parameters constrained by the experimental data point.

In order to solve for a parameter, the \texttt{name2} in the experimental data file (\texttt{totratios.data}) should be in the form \texttt{solveXXXX}, where \texttt{XXXX} is one of the fit parameters. The parameter limits set in \texttt{ratioset.data} still apply, every parameter solution outside of these limits is rejected. This helps rejecting unphysical solutions, such as $\lambda_s < 0$.

In principle, it is possible to solve for any parameter using any data point. However, many such combinations do not have a minimum, especially if the data point does not (or only marginally) depend on said parameter. If MINUIT takes a long time (e.g., many iterations) without converging to a minimum, there is a good chance that the minimization procedure will not work. It is thus advised to use this feature mainly to solve for the values of chemical potentials based on conservation laws. For instance, strangeness conservation can be assured in the system by solving for $\lambda_s$ by using the following line in the \texttt{totratios.data} file requiring net strangeness $\langle s-\bar{s}\rangle$ to vanish:\\[1mm]
\texttt{netstrang solvelams 0. 0. 0. 0}\\[1mm]
The baryochemical potential in a perfectly central Pb--Pb (Z=82, A=207) collision can be solved for using the baryon number and solving for $\lambda_q$:\\[1mm]
\texttt{netbaryon solvelamq 414. 0. 0. 0}\\[1mm]
and the corresponding charge conservation may be imposed on the fit with:\\[1mm]
\texttt{netcharge solvelmi3 164. 0. 0. 0}\\[1mm]
The \texttt{solve} statements have to come at the beginning of the experimental data file, otherwise the program will return an error.

An alternative to the exact solving for a parameter required by a conservation law is to require approximate conservation of a quantity. Treating a conservation law as a data point allows for detector acceptance corrections. A line such as:\\[1mm]
\texttt{netstrang totstrang  0.  0.01  0.  1}\\[1mm]
imposes strangeness conservation to within 1\%. We often impose charge over baryon number conservation with the line\\[1mm]
\texttt{netcharge netbaryon  0.39  0.02  0.  1}\\[-2mm]

The choice of implementing the conservation laws analytically or approximately (if at all) is left to the user. It is worth noting that exact solution is a more reliable procedure, however, it cannot be used very often considering the limited acceptance data from contemporary collider experiments.

\section{Running the program ---  \texttt{sharerun.data}}
\label{sec:fitting}
This file contains the instructions which are executed one by one during a program run.  Every line contains a separate operation, such as reading input files, assigning values to parameters, calculation of particle ratios, fitting model parameters to experimental data (i.e., minimizing $\chi^2$), plotting contours and $\chi^2$ profiles. The program will read one line at a time and execute the command until it reaches the end of the \texttt{sharerun.data} file. It is imperative that the user maintains the appropriate spacing of commands and values in this file, because this file is read as a formatted input, same as most of the input files described in previous sections. Any deviation from the expected number of characters may result in an unexpected behavior of the program ranging from a misinterpretation of a value to not recognizing the command at all and exiting prematurely with an error. Generally, there are two spaces between keywords and numbers. We specify the expected command format wherever necessary. Each command can be used multiple times with different input and output files. We shall now describe all the commands available to the user with a brief description.

\begin{figure}
\caption{\label{fig:sharerun.data}A typical \texttt{sharerun.data} file.}
\centering
\includegraphics[scale=0.95]{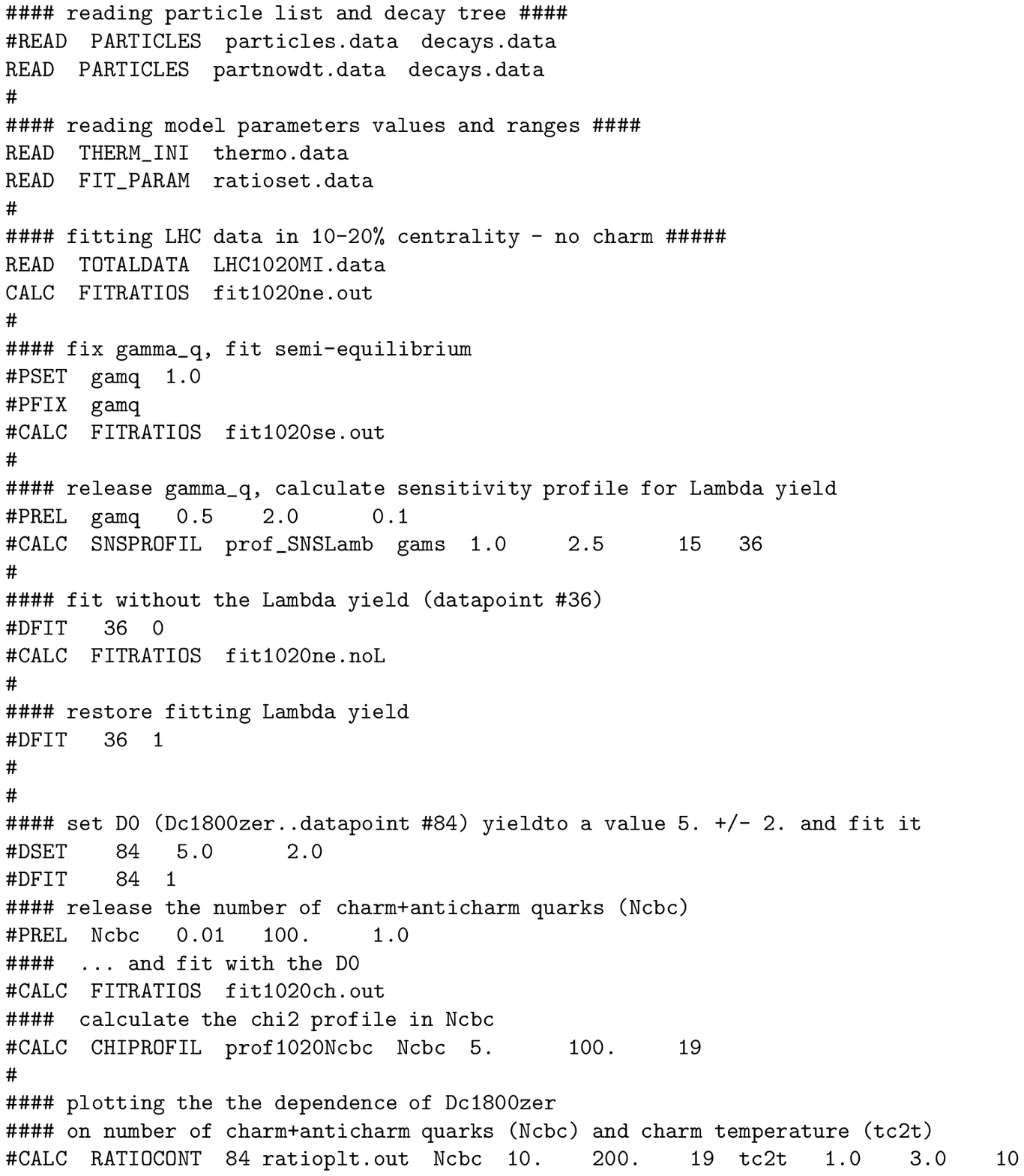}
\end{figure}

\subsection{Reading input files}
\begin{description}
\setlength{\itemsep}{-0.1cm}
\setlength{\labelwidth}{1.5cm}
\setlength{\itemindent}{0.5cm}
\item[\texttt{READ  THERM\_INI  $\langle$11--letter filename$\rangle$}]\ \\
Reads the specified file corresponding to the \texttt{thermo.data} file described in Section~\ref{sec:parameters}.
\item[\texttt{READ  FIT\_PARAM  $\langle$13--letter filename$\rangle$}]\ \\
Reads a file containing parameter ranges equivalent to the \texttt{ratioset.data} file described in seciton~\ref{sec:ratioset}.
\item[\texttt{READ PARTICLES $\langle$14--letter filename$\rangle$ $\langle$11--letter filename$\rangle$}]\ \\
Reads the file containing list of particle properties and a second file of corresponding decay tree. The files are considered being equivalent to \texttt{particles.data} and \texttt{decays.data} files described in Section~\ref{sec:particlelist} and~\ref{sec:decayfile-light}. 
\item[\texttt{READ TOTALDATA $\langle$14--letter filename$\rangle$}]\ \\
Reads experimental data file equivalent to the \texttt{totratios.data} file described in Section~\ref{sec:experimentaldata}.
\end{description}

\subsection{Parameter manipulation at run time}
\begin{description}
\setlength{\itemsep}{-0.1cm}
\setlength{\labelwidth}{1.5cm}
\setlength{\itemindent}{0.5cm}
\item[\texttt{PSET $\langle$4--letter tag$\rangle$ VALUE}]\ \\
The \texttt{PSET} command sets the thermodynamic parameter specified by the 4--letter tag to a \texttt{VALUE}. A real number is expected in \texttt{VALUE}, hence decimal point is necessary even with integer values. When the \texttt{thermo.data} file is read, a series of \texttt{PSET}-like commands is performed using all parameters and values read from the file.
\item[\texttt{PFIX $\langle$4--letter tag$\rangle$}]\ \\
Fixes the given model parameter to its current value. In all subsequent commands, this parameter will be kept fixed until released, or its value modified by another command. 

There are 5 additional \texttt{tag}s that are used to control the program options in connection with \texttt{PFIX}. The user can choose to use Boltzmann approximation and quantum (Fermi--Dirac and Bose--Einstein, as appropriate) statistics via the tags \texttt{bltz} and \texttt{fdbe} respectively. The default is quantum statistics. Furthermore, one can select to use the canonical ensemble treatment of the baryon number via \texttt{ce\_b} tag, and of strangeness via \texttt{ce\_s} tag. Selecting the program default grand canonical ensemble for both quantum numbers is done using the \texttt{gcal} tag.
\item[\texttt{PREL $\langle$4--letter tag$\rangle$ $\langle$lower limit$\rangle$ $\langle$upper limit$\rangle$ $\langle$step size$\rangle$ }]\ \\
Releases the given parameter and sets its range between the lower and upper limits and sets its initial step size. This command is equivalent to reading one line in the \texttt{ratioset.data} file. All three numbers in the \texttt{PREL} command are mandatory. Otherwise, zeros are assumed and any following calculation will produce unwanted results. The required format of this command is \texttt{(A4,2X,A4,3X,2F7.2,F9.5)}, i.e., after the parameter tag, the program expects in sequence; 3 spaces, 7 characters for parameter lower limit (counting the decimal point), 7 characters for the upper limit and 9 characters for the initial step size. Notice, that no spaces are required between the numbers, so the following line is valid:
\begin{verbatim}
PREL  temp   0.123450.23232000.00100
|     |      |      |      |        
      tag    Low    Upper  Step     
\end{verbatim}
This example releases the hadronization temperature parameter \texttt{temp} setting its limits to $T_{lower}=0.12345\,\mathrm{GeV}$ and $T_{upper}=0.23232\,\mathrm{GeV}$ and the initial step size in minimization to $0.001$. Similarly as in the above example, the user can take advantage of the precision allowed by the format specification of any command. It is, however, not needed very often.

\end{description}

\subsection{Datapoint run-time manipulation}
\noindent In some analysis, it may be beneficial to manipulate an experimental data point during the program run. The following two commands are used to accomplish this.
\begin{description}
\setlength{\itemsep}{-0.1cm}
\setlength{\labelwidth}{1.5cm}
\setlength{\itemindent}{0.5cm}
\item[\texttt{DFIT datapoint fit?}]\ \\
Switches if a data point on the \texttt{datapoint}-th line in the experimental data file will (\texttt{fit=1}) or will not (\texttt{fit=0}) be fitted during the evaluation of subsequent commands. The format of this command expects the keyword DFIT followed by 2 spaces and 5 characters allocated for each \texttt{datapoint} and \texttt{fit?}. The format is \texttt{(A4,2X,2I5)}.
\item[\texttt{DSET datapoint value error}]\ \\
Set the value of a \texttt{datapoint}-th ratio or quantity in the experimental data file to \texttt{value} and its uncertainty to \texttt{error}. The program expects the following format of the command; \texttt{(A4,2X,I5,2F10.5)}, i.e., 2 spaces after the keyword DSET, 5 characters allocated for the \texttt{datapoint} number, and 10 characters for each \texttt{value} and \texttt{error}.
\end{description}

\subsection{Calculation and fitting}
\begin{description}
\setlength{\itemsep}{-0.1cm}
\setlength{\labelwidth}{1.5cm}
\setlength{\itemindent}{0.5cm}
\item[\texttt{CALC RATIODATA $\langle$13--letter filename$\rangle$}]\ \\
Calculates the value of ratios and quantities read by the READ TOTALDATA command given the current values of thermal parameters (either read by the READ THERMO\_INI, or resulting from the last command). The output is written to a file with a 13--letter filename specified by the user. The output in the file has the general format of \texttt{RATIO name1/name2 $\langle$VALUE$\rangle$}. In case either of \texttt{name1} or \texttt{name2} is zero, zero is printed as a result of the ratio.
\item[\texttt{CALC RATIOPLOT $\langle$12--letter filename$\rangle$ $\langle$4--letter tag$\rangle$ L U P datapoint}]\ \\
This command calculates a ratio, yield or quantity specified on the \texttt{datapoint}-th line in the experimental data file as a function of the thermal parameter specified by the 4--letter tag. The program will vary the parameter from \texttt{L} to \texttt{H} in \texttt{P} equidistant steps and record in the output file the varying parameter values in the first and the property values in the second column. The output file can then be used as input to an external plotting program, such as GNUPlot, PAW, Xmgrace or Mongo. The format of this command is\\ \texttt{(A4,2X,A9,2X,A12,2X,A4,2X,2F5.1,2I5)}, i.e., two spaces between each word until the parameter tag, two spaces after the parameter tag, 5 characters reserved for each of the lower and upper limits followed by 5 characters for the number of points and last, 5 characters for the data point number.
\item[\texttt{CALC RATIOCONT datapoint $\langle$12--letter filename$\rangle$ $\langle$tag1$\rangle$ L1 U1 P1\dots}]\ \\
{\bf\texttt{\dots $\langle$tag2$\rangle$ L2 U2 P2}}\\ 
Calculates a ratio, yield or a thermodynamic quantity as a function of two parameters specified by \texttt{tag1} and \texttt{tag2} for parameter values ranging from \texttt{L1} to \texttt{U1} in \texttt{P1} steps for the first parameter and from \texttt{L2} to \texttt{U2} in \texttt{P2} steps for the second one. Similar to the above \texttt{RATIOPLOT} command, the data point is referred to by its line number (\texttt{datapoint}) in the experimental data file. (Note, that $100\times 100$ results in 10000 grid points to be calculated which may take a long time.) The general format of this command expected by the program is: (A4,2X,A9,I5,A12,2X,A4,2X,F5.1,2X,F5.1,2X,I3,2X,A4,2X,F5.1,2X,F5.1,2X,I3)\\
The output is written to a file with 13--letter filename specified by the user in a 3--column table and can be plotted by an external program capable of 3D plotting.
\item[\texttt{CALC  FITRATIOS $\langle$13--letter filename$\rangle$}]\ \\
This command minimizes $\chi^2/\mathrm{ndf}$ of the set of experimental data input by the last\\
\texttt{READ TOTALDATA} command varying parameters within ranges specified in\\
\texttt{READ FIT\_PARAM} starting with parameter initial values from \\
\texttt{READ THERM\_INI}. 


In our fitting procedure, we minimize the $\chi^2$ value function as function of SHM parameters. The $\chi^2$ function is the sum of squares of the relative difference between computed yields and experimental data.
\begin{equation}
\chi^2=\sum\limits_{i=1}^N \frac{(f_{i,\textrm{theory}}-f_{i,\textrm{experiment}})^2}{(\Delta f_{i,\textrm{systematic}}+\Delta f_{i,\textrm{statistical}})^2},
\end{equation}
where $f_i$ is the $i$--th investigated quantity with experimental error $\Delta f_i$ and $N$ is the number of data points. We seek the best fit to experimental particle yields and ratios using \texttt{MINUIT}~\citep{James:1975dr}, an optimization package part of the CERNLIB computational libraries. We evaluate the statistical significance (also called either $p$-value, or  confidence level -- CL) of our fits. CL is  defined as the probability that given a correct hypothesis (here the SHM model)  and Gaussian `noise' experimental errors (caution:  we are dealing with significant systematic data errors),  $\chi^2$ would in repeated measurements assume a value that is at or above the considered value -- clearly, the smaller $\chi^2$ is, the higher CL is, asymptotically approaching 100\%.   Values of statistical significance far below 50\% suggest that the model is not appropriate for the task of describing the experimental data. Regarding $\chi^2$ in the limit of many degrees of freedom $N_\mathrm{data}-N_\mathrm{parameter}=N_\mathrm{dof}\to\infty$, it is well known that $\chi^2/N_\mathrm{dof}= 1$ corresponds to CL of 50\%, see figure 36.2 in PDG~\cite{Beringer:1900zz}. For $N_\mathrm{dof} < 10$,  this figure shows that a considerably lower $\chi^2/N_\mathrm{dof}$ is necessary to reach CL of 50\%. For more complete discussion of significance, see PDG~\citep{Beringer:1900zz} section 36.2.2.

In case the minimization package MINUIT~\cite{James:1975dr} used by SHARE does not find a good reliable minimum via common  strategies, or the parameter errors cannot be reliably evaluated, MINOS subroutine is called to evaluate the parameter errors and potentially find a better minimum. This, unfortunately, significantly increases the computation time.

The output in the 13--letter file provided by the user has the general format:\\[2mm]
Header with time and date of minimization.\\[2mm]
Final thermal parameter values ( `+/-' error, when fitted),\\
followed by chemical potentials $\mu_B,\,\mu_S,\mu_{I_3}$\\
and phase space occupancies for each flavor $\gamma_u,\gamma_d,\gamma_s,\gamma_c$.\\[2mm]
Then, the detailed fit results are printed in a table format:\\
\texttt{TOP        BOTTOM        THEORY         EXP      ERROR        CHITERM    FEED-DOWN}\\[1mm]
where TOP refers to the denominator and BOTTOM to the numerator of the quantities defined in the input file (\texttt{totratios.data}), THEORY states the model value, EXP states the experimental value as given in the input file, ERROR is the combined statistical and systematic error, CHITERM is the $\chi$ contribution of the data point to the total $\chi^2/\mathrm{ndf}$ defined as:
\begin{equation}
\label{eq:chiterm}
\chi = \frac{f_{theory}-f_{experiment}}{\Delta f_{statistical} + \Delta f_{systematic}},\quad \text{( = 0 if not fitted ). }
\end{equation}
The $\chi$ is reported before squaring to keep the sign, i.e., information about the resulting model theoretical value being above or below the experimental data point. FEED-DOWN states the weak decay feed-down scheme or filename used for this particular data point. The end of the output file reports the number of degrees of freedom, total $\chi^2$, $\chi^2/\mathrm{ndf}$ and statistical significance of the fit.

At the end of each fit, two additional output files are created, \texttt{CharmFeedPrimary.data} and \texttt{CharmFeed.data}. The first one contains the primary yields of charm particles calculated by the CHARM module and the second one contains particle yields after charm decays, i.e., particle yields produced solely by charm. Note, that these two filenames are constant and the files get rewritten every time a fit is performed.

\item[\texttt{CALC FITNMINOS $\langle$13--letter filename$\rangle$}]\ \\
This command is equivalent to the above \texttt{FITRATIOS} regarding both format of the command line and function. The only difference is that parameter error evaluation MINOS~\cite{James:1975dr} is never called. Omitting the use of MINOS usually saves considerable amount of computing time, however, errors of the resulting thermal parameters will not be reliable and the minimum found has a higher chance of not being the global minimum.
\item[\texttt{CALC  PLOT\_DATA $\langle$3 13--letter filenames$\rangle$}]\ \\
Generates three files which are optimized to be graphed with an external 2D plotting package (GNUPlot, Xmgrace, PAW,\dots). The first file contains numerical list of ratios and quantities that were fitted, the second one contains a numerical list of ratios and quantities that were calculated and the last one will contain experimental values with errors. See the discussion in Section~\ref{sec:particleratios} on how to choose which file a quantity should be included in. The filenames on the command line should be separated by 2 spaces.
\item[\texttt{CALC  CHIPROFIL $\langle$12--letter filename$\rangle$ tag L U P} ] \ \\
This commands calculates a $\chi^2/\mathrm{ndf}$ profile of a parameter specified by the \texttt{tag}. The program divides the parameter range between \texttt{L}\,ower limit and \texttt{U}\,pper limit in \texttt{P} equidistant intervals, fixes the given parameter at the boundary of each interval (i.e., \texttt{P}+1 values including \texttt{L} and \texttt{U}) and fits other free parameters to the data as specified in the \texttt{ratioset.data} file or \texttt{PSET}/\texttt{PFIX}/\texttt{PREL} commands preceding this one. This procedure is equivalent to a sequence of \texttt{PSET} and \texttt{CALC  FITRATIOS} commands.

The main output of this command is the \texttt{filename} with \texttt{.prof} extension, which contains a 2--column table with the given parameter values in the first and the resulting $\chi^2/\mathrm{ndf}$ in the second column. The full output of all performed fits is stored in a log file with \texttt{\_\_log} extension. For each fitted parameter, a separate file is created containing 5 columns and has the parameter tag appended to its name. For instance, performing a temperature $\chi^2/\mathrm{ndf}$ profile with output file stored in `profTempLHC1', the values of $\gamma_q$ will be stored in `profTempLHC1\_gamq' file (provided that $\gamma_q$ is a free parameter of the fit).
The first column of each parameter file contains values of the \texttt{tag} parameter (temperature in the above example), the second column the fitted parameter values ($\gamma_q$ values in the above example), the third contains values of an associated bulk property (see Table~\ref{tab:associatedbulk} for a details), the fourth column states the $\chi^2/\mathrm{ndf}$ and the last column contains statistical significance.

The general format of this command is \texttt{(A4,2X,A9,2X,A12,2X,A4,2X,2F8.1,I5)}, in other words there has to be 2 spaces between the text strings up to the \texttt{tag}, after which there are 8 digits allocated for the \texttt{L}\,ower limit, 8 digits for the \texttt{U}\,pper limit and 5 digits for the number of \texttt{P}\,oints. 

\begin{table}
\caption{\label{tab:associatedbulk} Table of associated bulk properties with parameters printed in the profile output files.}
\begin{center}
\ttfamily\selectfont
\begin{tabular}{ll}
\hline
\textnormal{Parameter tag} & \textnormal{Associated bulk property}\\
\hline
temp  & totenergy prdensity\\
norm  & pressuret  prdensity\\
gamq  & entropy\_t prt\_yield \\
lamq  & netbaryon prt\_yield\\
mu\_b & netbaryon prt\_yield\\
gams  & totstrang prt\_yield\\
lams  & netstrang prt\_yield\\
mu\_s & netstrang prt\_yield\\
lmi3  & netcharge prt\_yield\\
mui3  & netcharge prt\_yield\\\hline
\end{tabular}
\end{center}
\end{table}
\item[\texttt{CALC SIGPROFIL $\langle$12--letter filename$\rangle$ tag L U P} ]\ \\
This command is very similar to the above \texttt{CHIPROFIL} in terms of both the command format and functionality. The only difference is that in the main output file \texttt{$\langle$filename$\rangle$.prof}, the statistical significance is printed in the second column rather than $\chi^2/\mathrm{ndf}$.
\item[\texttt{CALC FITPROFIL $\langle$12--letter filename$\rangle$ tag L U P datapoint} ]\ \\
This and the next two commands allow to study the fit parameter sensitivity to a particular data point in detail. The format of the command is the same as the one above, except for the extra integer \texttt{datapoint} identifier at the end (line number of the data point in the experimental data file). General format is \texttt{(A4,2X,A9,2X,A12,2X,A4,2X,2F8.1,2I5)}. The command produces a parameter \texttt{tag} profile for a fitted \texttt{datapoint} model prediction rather than the overall fit quality, as the previous two commands do. For example, if \texttt{datapoint} corresponds to $\pi/p$ ratio within the given data set, the program performs a $\chi^2$ profile calculation with respect to the parameter \texttt{tag} fixing it at \texttt{P} values  ranging from \texttt{L} to \texttt{U}.

The command creates several output files, namely \texttt{$\langle$filename$\rangle$.prof}, which contains the values of parameter \texttt{tag} in the first column and the $\chi$ term of the given \texttt{datapoint} defined by Eq.~\ref{eq:chiterm} in the second column. The output file with \texttt{.chi2} extension contains the overall $\chi^2/\mathrm{ndf}$ of the fits, and the extension \texttt{.stsg} represents statistical significance of the fits. For every fitted parameter other than \texttt{tag}, there is a correlation file with extension \texttt{\_$\langle$tag2$\rangle$} (e.g., \texttt{\_gamq}) created, which has the same format and contents as for the \texttt{CHIPROFIL} command above. Full output for all the fits is stored in a log file with extension \texttt{\_\_log}.
\item[\texttt{CALC DATPROFIL $\langle$12--letter filename$\rangle$ tag L U P datapoint}]\ \\
This command has the same command line structure, functionality and output files as the \texttt{FITPROFIL} command above, except the values of the \texttt{datapoint} (rather than its $\chi$ term contribution to the overall $\chi^2$ of the fits) is printed in the output file \texttt{$\langle$filename$\rangle$.prof}.
\item[\texttt{CALC SNSPROFIL $\langle$12--letter filename$\rangle$ tag L U P datapoint} ]\ \\
Same as the above two commands, except the output file \texttt{$\langle$filename$\rangle$.prof} now contains the \emph{sensitivity} of the given \texttt{datapoint} to the particular parameter \texttt{tag}. The sensitivity is defined as a ratio of the data point's SHM prediction for a given parameter \texttt{tag} value to the SHM prediction of the \texttt{datapoint} for the best fit value of that parameter.
\item[\texttt{CALC CHI2\_CONT $\langle$9--letter filename$\rangle$ deviation tag1 tag2} ]\ \\
This command calculates the $\chi^2/\mathrm{ndf}$ contour for the parameters \texttt{tag1} and \texttt{tag2}. The program will output to the file \texttt{$\langle$filename$\rangle$} pairs of parameter \texttt{tag1} and \texttt{tag2} values that correspond to $\chi^2/\mathrm{ndf}=\mathtt{deviation}\,\times\,(\chi^2/\mathrm{ndf})_{\mathrm{bestfit}}$. The general format of the command is \texttt{(A4,2X,A9,2X,A9,1X,F4.1,2X,A4,2X,A4)}
\end{description}

\subsection{Run log (\texttt{sharerun.out})}
The complete `log' for each run is saved in a file \texttt{sharerun.out}. This includes:
\begin{itemize}
\setlength{\itemsep}{-2mm}
\setlength{\itemindent}{5mm}
\item The contents of each input file as read in by the program (useful for input file format check)
\item A list of performed operations.
\item MINUIT output. Parameter correlation matrix can be found here, for example.
\item The content of each output file (in the same format as in the output file)
\end{itemize}
If the program does not encounter any problems during the run, a message:\\ \texttt{RUN TERMINATED SUCCESSFULLY},\\is printed on screen and at the end of the \texttt{sharerun.out} file. In case an error occurs during a run, the program reports it to the user on screen. More information about the error is reported in the \texttt{sharerun.out} file.

\subsection{Fitting pitfalls, parameter sensitivity to data}
\label{sec:sensitivity}
Convergence to the very same set of best  parameters cannot be expected when one or more of the parameters is  insensitive to any of the fitted experimental data points. In this case, the minimum of $\chi^2$ function is a domain in the respective parameter space, and a point in this domain is chosen in a  quasi-random manner. As an example of this situation, consider the data from LHC.  When fitting particle yields and ratios, one can constrain the values of chemical non-equilibrium parameters $\gamma_q$ and $\gamma_s$ along with $T, V$, as these are constrained by the precisely measured  numerous baryon and meson yields~\cite{Petran:2013lja}. However, in the LHC environment, it is challenging to constrain the value of baryochemical potential $\mu_B$. This is especially the case when fitting with charm. While a measurement of any charm hadron yield, for instance $D^0$ meson, with 10\% precision constrains the charm yield to a narrow range in $N_{c\bar{c}}$, the charm feed-down introduces additional uncertainty in the value of $\mu_B$ due to charm baryon production and decays. Moreover, feed-down impacts hadron yields and hence the other fitted parameters adjust accordingly. In order to measure  the  baryochemical potential $\mu_B$, one would need, for example, a relative baryon--antibaryon  difference measurement, $(\overline{B_i}-B_i)/(\overline{B_i}+B_i)$ for a baryon $B_i$ obtained with a precision comparable to the other input data.

We thus advise fixing insensitive parameters to most likely values, especially when fitting charm at LHC. Therefore, until additional data fixing $\mu_B$ becomes available, one should proceed with zero baryochemical potential, that is fix $\lambda_s=\lambda_q=1$, in the exploration of charm effect on hadronization. 

\section{Example: fit with prescribed  charm yield and bulk energy density}
\label{sec:example_output}
\begin{figure}
\centering
\includegraphics[width=\columnwidth]{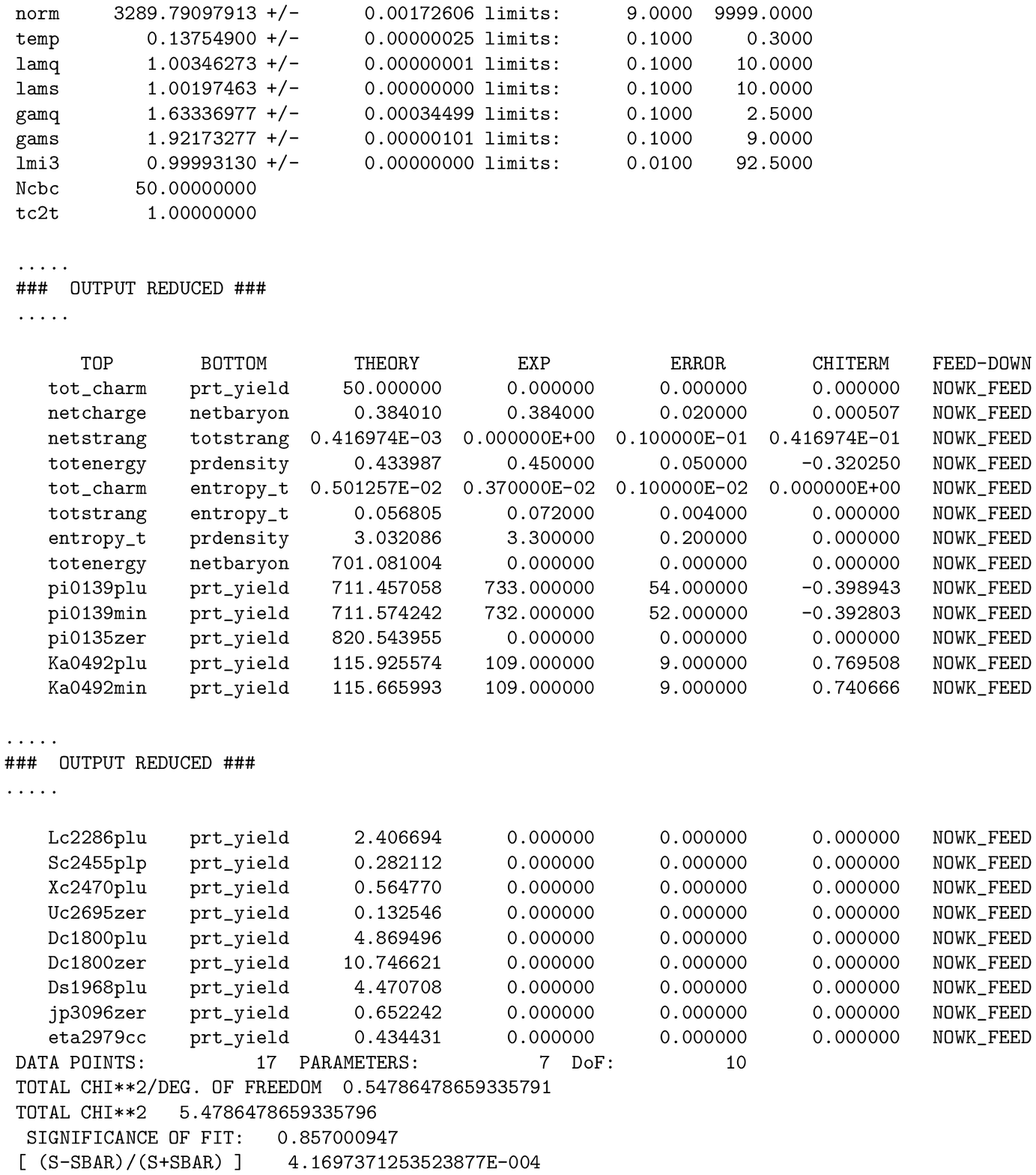}
\caption{\label{fig:charm_output} Sample of \textit{SHARE with CHARM} output for a prescribed 50 $c\!+\!\bar{c}$ quarks at hadronization. Non-charm most central (0-5\%) LHC data used in~\cite{Petran:2013lja} is fitted, charm hadron decays are injected into hadron multiplicities. We also prescribe the bulk energy density $d\varepsilon/dy=0.45\pm0.05\,\mathrm{GeV/fm}^3$. Note that the fit outcome is 85\% confidence level for 10 degrees of freedom. (Program output was reduced to fit in the figure.)}
\end{figure}

We demonstrate the program capabilities by showing in Figure~\ref{fig:charm_output} a fit to LHC data we already characterized recently~\cite{Petran:2013lja}. This data does not yet comprise any charm hadron yields and thus we cannot fit here the charm yield present. The new charm module capabilities are demonstrated by adding an ad-hoc yield of charm at hadronization, $N_{c\bar{c}} = 50$ charm plus anti-charm quarks (25 $c\bar{c}$ pairs). In order to demonstrate how physical bulk properties can be used in the fit, we require a specific hadronization conditions by fitting a bulk physical property, in this case we decided to illustrate the example by prescribing the energy density $\varepsilon=0.45\pm 0.05\,\mathrm{GeV/fm}^3$.

We show in Figure~\ref{fig:charm_output} a reduced program output with final statistical parameters ($T,\gamma_q,\gamma_s,\dots$), fitted physical properties of the bulk, followed by charm baryon, charm meson and charmonium yields. The bottom of the output shows the fit quality in terms of total $\chi^2$, reduced $\chi^2/\mathrm{ndf}$, and statistical significance $\sim 85\%$. The main difference between the here presented fit and the `standard' fits  is that nearly 50 charm hadron decays producing as many as 200 non-charm hadrons containing strangness and multi-strangeness. As the Figure~\ref{fig:charm_output} shows the fit is very good and converges to a relatively low temperature of $T=137.5$~MeV. The reduction of temperature by a few MeV is due to the injection of charm hadron decay products into the fitted hadron abundances.

Further  LHC physics related  use of the program is presented in  our recent publications~\cite{Petran:2013lja,Petran:2013kpa,Petran:2013qla}.

\section{Installation}
\label{sec:installation}
\subsection{Pre-requisites}
Before you start installing SHARE with CHARM, make sure you have the following installed on your system:
\begin{itemize}
\setlength{\itemsep}{-2mm}
\setlength{\itemindent}{5mm}
\item Fortran 77 compiler, we have tested GNU \texttt{gfortran} and Intel \texttt{ifort} compilers.
\item C++ compiler (including C++ standard libraries), we have tested GNU \texttt{g++} and Intel \texttt{icpc} compilers.
\item CERNLIB libraries. They are available in standard repositories for most GNU Linux distributions, alternatively, the source code is available for download at~\cite{cernlib}. Debian based linux distribution (Ubuntu,\dots) users should locate and install the package \texttt{cernlib} from their system repository. Red Hat base (Fedora Project, SLC,\dots) can install cernlib from a rpm package available, for example, at~\cite{cernlib-rpm}.
\end{itemize}

\subsection{Organization of the program package}
The SHARE with CHARM program comes in a single .zip archive labeled \texttt{sharev3.zip}. The package is available at {\it http://www.physics.arizona.edu/\textasciitilde{}gtshare/SHARE/share.html}.\\
After obtaining the program package, unpack it using the following command (it will create a folder called \texttt{sharev3} with all the files in it):\\[2mm]
{\bf \texttt{unzip sharev3.zip}}\\

The following files will be created in this directory from the archive contents. The files are enough for a `representative' run of SHARE with CHARM.
\begin{description}
\setlength{\itemsep}{-2mm}
\setlength{\labelwidth}{1.5cm}
\setlength{\itemindent}{5mm}
\item[\texttt{src/sharev3.0.0.F}] SHARE source code in Fortran 77.
\item[\texttt{src/CharmDistribution\_v1.9.cpp}] CHARM module source code in C++.

\item[\texttt{decays.data}] The complete Particle Data Group decay tree of light hadrons (Section~\ref{sec:decayfile-light}).
\item[\texttt{HFfeed.data}] The complete Particle Data Group decay tree of charm hadrons (Section~\ref{sec:HFfeed}).
\item[\texttt{LHC1020MI.data}] A representative input data files (Section~\ref{sec:experimentaldata}) containing experimental yields data for centrality bin 10-20\% based on data from the ALICE experiment as of August 2013, see~\cite{Petran:2013lja} and references therein.
\item[\texttt{makefile}] A makefile of the project, useful for program compilation, see Section~\ref{sec:make-compilation} below.
\item[\texttt{particles.data}] Particle properties with full widths (Section~\ref{sec:particlelist}).
\item[\texttt{partnowdt.data}] Particle properties with no widths (Section~\ref{sec:particlelist}). Calculations with this input file require significantly less computational time, and the use of full widths has not yet been justified, see for example recent comparison in~\cite{Petran:2013dea}.
\item[\texttt{ratioset.data}] The parameter ranges input file (Section~\ref{sec:ratioset}).
\item[\texttt{ratioset.test}] The parameter ranges input file for Quick Start test run.
\item[\texttt{sampleoutput}] Directory containing sample output files resulting from a `representative' sample run with input files we provide as part of the package.
\item[\texttt{sharerun.data}] A `representative' run input file (Section~\ref{sec:fitting}) including an analysis of hadron yields at LHC presented by members of our collaboration in~\cite{Petran:2013lja}.
\item[\texttt{thermo.data}] The list of model parameters (Section~\ref{sec:parameters}).
\item[\texttt{weak.feed}] A sample weak decay file (Section~\ref{sec:weakdecays}).

\end{description}

\subsection{Recommended compilation --- using \texttt{make}}
\label{sec:make-compilation}
The recommended way to compile SHARE with CHARM is to make use of the GNU \texttt{make} utility~\cite{make}, available on most GNU Linux systems (if this is not your case, proceed to Section~\ref{sec:manualcompilation}). The most common combination of available compilers is \texttt{gfortran} and \texttt{g++}. If this is your case, after unzipping the contents of the program package into a folder, use the command 

{\bf\texttt{make}}

\noindent in this directory to compile the program. If you are using different compilers, edit the header of the included \texttt{makefile} to specify compilers available on your system. When done, use \texttt{make} (as above) to compile SHARE with CHARM.

Using \texttt{make} with the included makefile processes the two source files in the \texttt{src} folder, compiles each of them into an object file, which is stored in the \texttt{obj} folder and then links the object files together with CERN libraries into an executable \texttt{share}, which is ready to be run from the command line. We provide the two most common targets in the makefile, \texttt{all} to compile the program and \texttt{clean} to remove objects and executable file.
You may also need to specify the location of the CERN libraries on your system in the \texttt{makefile}, for instance,\\ \texttt{LIBS = -L/usr/local/cernlib/2006/lib -lstdc++ \dots}

\subsection{Manual compilation}
\label{sec:manualcompilation}
SHARE with CHARM consists of two source code files, both located in the \texttt{src} subdirectory. Manual compilation consists of three steps. 
\begin{enumerate}[leftmargin=*]
\item Compiling SHARE Fortran code into an object. For example, using the GNU Fortran compiler \texttt{gfortran}, this is done via the following command (in the \texttt{src} folder):\\
{\bf\texttt{gfortran sharev3.0.0.F -c -o sharev3.0.0.o}}\\
Note the \texttt{-c} option to prevent linking. You may need to consult your Fortran compiler documentation for equivalent command line options.
\item Compiling the CHARM module written in C++ is accomplished similarly to the above step, using an example command line with GNU C++ compiler \texttt{g++}:\\
{\bf\texttt{g++ CharmDistribution\_v1.9.cpp -c -o CharmDistribution\_v1.9.o}}
\item Linking all object files with the necessary libraries into an executable binary file \texttt{share} is accomplished using the Fortran compiler used in step 1 on the following command line :\\
{\bf\texttt{gfortran sharev3.0.0.o CharmDistribution\_v1.9.o \textbackslash \\ \, \hspace*{5mm} -lstdc++ -lkernlib -lmathlib -lpacklib -o share}}\\
In case you compiled CERN libraries manually, you may need to specify the location of the libraries on the last command line using the \texttt{-L/<cernlib location>/lib} option in order to link them properly. If necessary, consult your compiler and CERN libraries documentation for details.
\end{enumerate} 

When all three steps are completed without errors, move (copy) the executable binary of SHARE with CHARM (\texttt{share} in the above example) to the parent directory (i.e., one directory up) where all SHARE input files are located. The program is invoked with \texttt{./share} command (unless a different name was chosen by the user during compilation). The program opens the \texttt{sharerun.data} file located in the current directory and performs tasks specified therein. See Section~\ref{sec:fitting} for details on how to run the program. The provided copy of \texttt{sharerun.data} should produce detailed output showing the program capabilities, which can used to check correct program operation by comparing with the provided sample output in the \texttt{sampleoutput} directory. Tests were run on both 32 and 64 bit processors with two different versions of CERN libraries. Differences between platforms appear when fitted parameters are not constrained by data, see Section~\ref{sec:sensitivity}.

\section{Summary of current SHM status}
SHARE with CHARM is an analysis tool developed specifically to study particle production in relativistic heavy-ion collisions spanning an energy range from compact baryonic matter through the entire RHIC range up to top LHC energy. SHARE with CHARM is particularly suitable to address the following questions (italic font show new features of this release):
\begin{itemize}
\setlength{\itemsep}{0mm}
\setlength{\parskip}{0mm}
\item What is chemical freeze-out temperature, chemical potentials, and volume?
\item Which quark flavors are in chemical equilibrium at hadronization and those that are not, how abundant are they?
\item What are the physical bulk properties of the hadronizing fireball?
\item Are particle yield fluctuations compatible with hadron yields and ratios?
\item \textit{Is charm subject to statistical hadronization?} 
\item \textit{How large is the contribution of charm hadron decays to  the final light hadron yields?} 
\item \textit{How does accounting for charm decay feed change the  hadronization conditions?}
\item \textit{Does charm hadronize at the same temperature as the light hadron  freeze-out?}
\end{itemize}

The need for the new version of SHARE with CHARM arises from the necessity of including charm hadrons into statistical hadronization model as they become significant in heavy-ion collision experiments at LHC energy range. In the upgraded program, a flexible treatment of charm hadrons has been introduced, and we provide  full charm hadron list and, more importantly, a full decay tree compiled to the best of current knowledge of charm decays in a procedure that assures cross-particle symmetries and consistence.

At the time of preparing this publication, there is no charm hadron yield measured at LHC. With partial measurement of the $D^0$--$p_T$-spectrum, we estimate its invariant yield $dN/dy$ to be in the range of $1.3 < dN_{D^0}/dy < 9.0$, corresponding to $N_{c\bar{c}} \in (6,45)$. However, in primary parton collisions a more generous result comes from  scaling the total charm cross-section in $pp$ collisions $\sigma_{cc}$, which implies $N_{c\bar{c}} = 246\pm154$~\cite{Nelson:2012bc}, where most of the uncertainty is inherent from the $\sigma_{cc}$ uncertainty. SHARE with CHARM is capable of exploring both of these regions. As soon as a single charm hadron yield data becomes available, fitting this hadron using  SHARE with CHARM will help to constrain the total amount of charm present at hadronization allowing to cross check with the production models. With a second charm hadron yield, the difference between light and charm hadronization temperature can  be constrained. And finally, with additional charm particle yields, we will be able to answer if charm hadronizes according to statistical hadronization principles.

We believe that at the time of writing, SHARE with CHARM is the only SHM implementation capable of:
\begin{itemize}
\setlength{\itemsep}{0mm}
\setlength{\parskip}{0mm}
\item accounting for full chemical non-equilibrium of all four quark flavors, $u,d,s$ and $c$,
\item characterizing and/or fitting bulk properties of the particle source,
\item evaluating the produced hardon fluctuations,
\item quantifying the charm hadron contribution to hadron abundances,
\end{itemize}
aside from providing the other features necessary to describe soft hadron production in relativistic heavy-ion collisions in the entire energy range.

\section*{Acknowledgement}
\begin{sloppypar}
Work supported by  the U.S. Department of Energy, Grants no. DE-FG02-04ER41318 (MP, JL, JR) and DE-FG02-93ER40764 (GT). Laboratoire de Physique Th{\' e}orique et Hautes Energies, LPTHE, at University Paris 6 is supported by CNRS as Unit{\' e} Mixte de Recherche, UMR7589. GT acknowledges  financial support received from the Helmholtz International Centre for FAIR within the framework of the LOEWE program (Landesoffensive zur Entwicklung Wissenschaftlich\--\"{o}konomischer Exzellenz) launched by the State of Hesse.
\end{sloppypar}



\begin{thebibliography}{0}
\bibitem{Beringer:1900zz-front} 
  J.~Beringer {\it et al.}  [Particle Data Group Collaboration],
  Phys.\ Rev.\ D {\bf 86}, 010001 (2012).


\bibitem{Torrieri:2004zz-front} 
  G.~Torrieri, S.~Steinke, W.~Broniowski, W.~Florkowski, J.~Letessier and J.~Rafelski,
  Comput.\ Phys.\ Commun.\  {\bf 167}, 229 (2005)
  [nucl-th/0404083].


\bibitem{Torrieri:2006xi-front} 
  G.~Torrieri, S.~Jeon, J.~Letessier and J.~Rafelski,
  Comput.\ Phys.\ Commun.\  {\bf 175}, 635 (2006)
  [nucl-th/0603026].
  
  \end{thebibliography}

\begin{thebibliography}{00}

\bibitem{Rafelski:2003zz} 
  J.~Kapusta, B.~Muller and J.~Rafelski,
  ``Quark-Gluon Plasma: Theoretical Foundations'',
  Elsevier Science, (2003).

\bibitem{Torrieri:2004zz} 
  G.~Torrieri, S.~Steinke, W.~Broniowski, W.~Florkowski, J.~Letessier and J.~Rafelski,
  Comput.\ Phys.\ Commun.\  {\bf 167}, 229 (2005).


\bibitem{Letessier:2002gp} 
  J.~Letessier and J.~Rafelski,
  ``Hadrons and quark--gluon plasma'',
  Camb.\ Monogr.\ Part.\ Phys.\ Nucl.\ Phys.\ Cosmol.\  {\bf 18}, 1 (2002).

\bibitem{Koch:1986ud} 
  P.~Koch, B.~Muller and J.~Rafelski,
  Phys.\ Rept.\  {\bf 142}, 167 (1986).



\bibitem{Beringer:1900zz} 
  J.~Beringer {\it et al.}  [Particle Data Group Collaboration],
  Phys.\ Rev.\ D {\bf 86}, 010001 (2012).


\bibitem{Petran:2011aa} 
  M.~Petran, J.~Letessier, V.~Petracek and J.~Rafelski,
  Acta Phys.\ Polon.\ Supp.\  {\bf 5}, 255 (2012).


\bibitem{Matsui:1986dk} 
  T.~Matsui and H.~Satz,
  Phys.\ Lett.\ B {\bf 178}, 416 (1986).

\bibitem{Gazdzicki:1999rk} 
  M.~Gazdzicki and M.~I.~Gorenstein,
  Phys.\ Rev.\ Lett.\  {\bf 83}, 4009 (1999).

\bibitem{Andronic:2003zv} 
  A.~Andronic, P.~Braun-Munzinger, K.~Redlich and J.~Stachel,
  Phys.\ Lett.\ B {\bf 571}, 36 (2003).

\bibitem{Andronic:2008gm} 
  A.~Andronic, P.~Braun-Munzinger, K.~Redlich and J.~Stachel,
  J.\ Phys.\ G {\bf 35}, 104155 (2008).

\bibitem{Schroedter:2001rh} 
  M.~Schroedter, R.~L.~Thews and J.~Rafelski,
  J.\ Phys.\ G {\bf 27}, 691 (2001).


\bibitem{Rafelski:1996hf}
  J.~Rafelski, J.~Letessier and A.~Tounsi,
  Acta Phys.\ Polon.\ B {\bf 27}, 1037 (1996).


\bibitem{Schroedter:2000ek} 
  M.~Schroedter, R.~L.~Thews and J.~Rafelski,
  Phys.\ Rev.\ C {\bf 62}, 024905 (2000)


\bibitem{BraunMunzinger:2000px} 
  P.~Braun-Munzinger and J.~Stachel,
  Phys.\ Lett.\ B {\bf 490}, 196 (2000).

\bibitem{Thews:2000rj} 
  R.~L.~Thews, M.~Schroedter and J.~Rafelski,
  Phys.\ Rev.\ C {\bf 63}, 054905 (2001).


\bibitem{Gorenstein:2000ck} 
  M.~I.~Gorenstein, A.~P.~Kostyuk, H.~Stoecker and W.~Greiner,
  Phys.\ Lett.\ B {\bf 509}, 277 (2001).

\bibitem{Kostyuk:2003kt} 
  A.~P.~Kostyuk, M.~I.~Gorenstein, H.~Stoecker and W.~Greiner,
  Phys.\ Rev.\ C {\bf 68}, 041902 (2003)

 


\bibitem{Li:2007xf} 
  J.~-W.~Li and D.~-S.~Du,
  Phys.\ Rev.\ D {\bf 78}, 074030 (2008).



\bibitem{Kuznetsova:2006bh} 
  I.~Kuznetsova and J.~Rafelski,
  Eur.\ Phys.\ J.\ C {\bf 51}, 113 (2007).

 
\bibitem{Nelson:2012bc} 
  R.~E.~Nelson, R.~Vogt and A.~D.~Frawley,
  Phys.\ Rev.\ C {\bf 87}, 014908 (2013).
  and private communication with R.~Vogt


\bibitem{Torrieri:2005va} 
  G.~Torrieri, S.~Jeon and J.~Rafelski,
  Phys.\ Rev.\ C {\bf 74}, 024901 (2006).

\bibitem{Pruneau:2002yf} 
  C.~Pruneau, S.~Gavin and S.~Voloshin,
  Phys.\ Rev.\ C {\bf 66}, 044904 (2002).

\bibitem{F77standard}
  Fortran 77 Standard,
  {\it http://www.fortran.com/F77\_std/rjcnf0001.html}, Accessed 20 Sept 2013, (or any F77 reference manual).

\bibitem{Petran:2013dea} 
  M.~Petran, J.~Letessier, V.~Petracek and J.~Rafelski,
  arXiv:1309.6382 [hep-ph], in proceedings of Strangeness in Quark Matter 2013, 22-27 July 2013, Birmingham, UK.

\bibitem{Torrieri:2006xi} 
  G.~Torrieri, S.~Jeon, J.~Letessier and J.~Rafelski,
  Comput.\ Phys.\ Commun.\  {\bf 175}, 635 (2006).

\bibitem{James:1975dr} 
  F.~James and M.~Roos,
  Comput.\ Phys.\ Commun.\  {\bf 10}, 343 (1975).

\bibitem{Petran:2013lja} 
  M.~Petran, J.~Letessier, V.~Petracek and J.~Rafelski,
  Phys.\ Rev.\ C {\bf 88}, 034907 (2013).

\bibitem{Petran:2013kpa}
  M.~Petráň, J.~Letessier, V.~Petráček and J.~Rafelski,
  arXiv:1310.2551 [hep-ph],Strangeness in Quark Matter 2013 proceedings in press.

\bibitem{Petran:2013qla}
  M.~Petran and J.~Rafelski,
  Phys.\ Rev.\ C {\bf 88} (2013) 021901
  [arXiv:1303.0913 [hep-ph]].


\bibitem{cernlib}
  Homepage of CERNLIB,
  {\it http://cernlib.web.cern.ch/cernlib/}, Accessed 30 Sept 2013.


\bibitem{cernlib-rpm}
	CERNLIB rpm package list,
 {\it http://rpmfind.net/linux/rpm2html/search.php?query=cernlib}, Accessed 6 Oct 2013.



\bibitem{make}
  Homepage of the `make' utility,
  {\it http://www.gnu.org/software/make/}, Accessed 30 Sept 2013.







\end{thebibliography}
\end{document}